\documentclass[12pt, a4paper]{article}
            
\usepackage{a4}
\usepackage{graphicx}
\usepackage{amsmath}
\usepackage{amssymb}
\usepackage{subfig}
\usepackage[all]{xy}  

\usepackage{hyperref}
\hypersetup{pdftitle={Cohomology of Line Bundles: Applications}, 
      pdfauthor={Ralph Blumenhagen, Benjamin Jurke, Thorsten Rahn, Helmut Roschy}, 
      pdfsubject={Algebraic Geometry, Mathematical Methods in Physics},
      pdfstartview={FitH}, pdfpagelayout={TwoColumnRight},
      bookmarksopen, bookmarksnumbered, bookmarksopenlevel=2}

\newcommand{\cohomCalg}{{\text{\fontfamily{put}\bfseries\footnotesize\selectfont cohomCalg}}}
\newcommand{\CPP}{C\nolinebreak\hspace{-.05em}\raisebox{.4ex}{\tiny\bf +}\nolinebreak\hspace{-.10em}\raisebox{.4ex}{\tiny\bf +}}

\newcommand{\ttmat}[4]{\Big(\!\!\begin{array}{cc} #1 & #2 \\ #3 & #4 \end{array}\!\!\Big)}

\newcommand{\beq}{\begin{equation}}  \newcommand{\eeq}{\end{equation}}
\newcommand{\bal}{\begin{aligned}}   \newcommand{\eal}{\end{aligned}}

\newcommand{\IMAT}{1\!\! 1}

\def\IR{\mathbb{R}}
\def\IC{\mathbb{C}}
\def\IP{\mathbb{P}}
\def\IZ{\mathbb{Z}}

\def\cN{\mathcal{N}}
\def\cO{\mathcal{O}}
\def\cS{\mathcal{S}}
\def\cQ{\mathcal{Q}}
\def\cE{\mathcal{E}}
\def\cI{\mathcal{I}}

\def\fh{\mathfrak{h}}
\def\fC{\mathfrak{C}}

\def\fhh{\mathfrak{h}}

\def\clap#1{\hbox to 0pt{\hss#1\hss}}
\def\mllap{\mathpalette\mathllapinternal}

\def\mclap{\mathpalette\mathclapinternal}
\def\mathllapinternal#1#2{%
\llap{$\mathsurround=0pt#1{#2}$}}

\def\mathclapinternal#1#2{%
\clap{$\mathsurround=0pt#1{#2}$}}

\def\ce{\mathrel{\mathop:}=}  

\def\fto{\longrightarrow}
\def\injto{\lhook\joinrel\relbar\!\!\:\joinrel\rightarrow}
\def\surjto{\relbar\joinrel\twoheadrightarrow}

\def\uspc{${}^\big.$}   
\def\lspc{${}_\big.$}

\begin{document}

\baselineskip=14pt
\parskip 5pt plus 1pt 

\vspace*{-1.5cm}
\begin{flushright}    
  {\small
  MPP-2010-134 \\
  NSF-KITP-10-131}
\end{flushright}

\vspace{2cm}
\begin{center}        
  {\LARGE
  Cohomology of Line Bundles:\\[0.2cm]
  Applications
  }
\end{center}

\vspace{0.75cm}
\begin{center}        
  Ralph Blumenhagen${}^{1,2}$, Benjamin Jurke${}^{1,2}$, Thorsten Rahn${}^{1}$, Helmut Roschy${}^{1}$
\end{center}

\vspace{0.15cm}
\begin{center}        
  \emph{${}^{1}$Max-Planck-Institut f\"ur Physik, F\"ohringer Ring 6, \\ 
               80805 M\"unchen, Germany} \\[5mm]
  \emph{${}^{2}$Kavli Institute for Theoretical Physics, Kohn Hall, UCSB, \\ 
               Santa Barbara, CA 93106, USA}
\end{center} 

\vspace{2cm}


\begin{abstract}
Massless modes of both heterotic and Type~II string compactifications on compact manifolds are determined by vector bundle valued cohomology classes. Various applications of our recent algorithm for the computation of line bundle valued cohomology classes over toric varieties are presented. For the heterotic string, the prime examples are so-called monad constructions on Calabi-Yau manifolds. In the context of Type~II orientifolds, one often needs to compute equivariant cohomology for line bundles, necessitating us to generalize our algorithm to this case. Moreover, we exemplify that the different terms in Batyrev's formula and its generalizations can be given a one-to-one cohomological interpretation. 
\end{abstract}

\clearpage

\newpage


\tableofcontents

\newpage
\section{Introduction}

Since the mid-eighties \cite{Candelas:1985en} string compactifications to four space-time dimensions on compact Calabi-Yau (CY) manifolds have been under intense investigation. This has led to not only a better understanding of the space of possible string models, these days called the string landscape, but also to new developments in mathematics, such as for instance mirror symmetry. Of particular interest are string models with four-dimensional $\cN$=1 space-time supersymmetry, a realistic gauge group and chiral matter transforming in certain representations of the gauge group. Many classes have been considered, such as heterotic strings on Calabi-Yau manifolds with the bosonic, left-moving sector coupling to a vector bundle on the Calabi-Yau or Type~IIB orientifolds with intersecting D7-branes, which for chirality carry non-trivial line bundles (see the review \cite{Blumenhagen:2006ci} and refs.~therein). These latter ones are related by the Sen-limit to F-theory compactifications on Calabi-Yau fourfolds, which have recently been studied intensively \cite{Donagi:2008ca, Beasley:2008dc}. 

For all these compactifications, one is particularly interested in the massless excitations of the string, which in the large volume regime are determined by the zero modes of the Dirac respectively Laplace operator on the curved manifold. These modes are counted by certain cohomology classes over (submanifolds of) the Calabi-Yau manifold. Often it is not so hard to compute at least the chiral massless spectrum by an index theorem, but the complete computation, including vector-like matter states, involves more sophisticated methods. 

A large class of Calabi-Yau manifolds is given by complete intersections of hypersurfaces in ambient toric varieties. In this case one can make use of the fact that for toric spaces many combinatorial tools are available for the computation of topological quantities like the intersection form, Chern classes etc. Moreover, here one can naturally define vector bundles by certain (exact) sequences involving just line bundles, as for instance the monad construction (see e.g.~\cite{Witten:1993yc, Distler:1996tj, Anderson:2007nc, Anderson:2008uw}) or so-called extensions (see e.g.~\cite{Braun:2005zv}). For such bundles, the computation of the bundle valued cohomology can be traced back to the evaluation of line bundle valued cohomology classes over the toric ambient space. Based on earlier partial results \cite{Distler:1996tj, Blumenhagen:1997vt, Blumenhagen:1997cn}, in \cite{Blumenhagen:2010pv} we have conjectured a new algorithm for their determination and have developed a high-performance C/{\CPP} implementation \cite{cohomCalg:Implementation} of it. This conjecture was then proven  shortly afterwards in \cite{Jow:2010arXiv1006.0780J, CohomOfLineBundles:Proof}.\\
 
This paper can be considered as the third in the row of \cite{Blumenhagen:2010pv, CohomOfLineBundles:Proof} and shows the many possible string theoretic applications of the algorithm. Here we are not heading for  new kinds of realistic string models, but merely collect the mathematical tools for the determination of the massless matter spectrum and show how our algorithm helps tremendously in their actual computation. In section~\ref{sec_preliminaries}, we first review the algorithm and apply it to the determination of the Hodge numbers for toric varieties. This is generalized in section~\ref{sec_calabiyau} to vector bundle valued cohomology classes over hypersurfaces and complete intersections in toric varieties. The essential tool here is the Koszul sequence, which allows to uplift the cohomology over a submanifold to the cohomology over the ambient space. We discuss both the general problem and a couple of concrete examples.

Section~\ref{sec_orientifolds} deals with the appearing topological questions in orientifold constructions, where often just knowing the cohomology is not sufficient, but one also needs information how it transforms under the $\IZ_2$ orientifold projection. In mathematical terms, one needs to compute $\IZ_2$ equivariant cohomology classes. We will show that our algorithm is sort of tailor-made to be easily generalized to equivariant cohomology classes. This is due to the fact that we have a concrete representative for each element so that we can directly determine the orientifold action on it. This extended algorithm is tested by a couple of non-trivial examples. 

Finally, section~\ref{sec_genbatyrev} contains a study of the different contributions to the combinatorial Batyrev formula for the Hodge numbers of a Calabi-Yau manifold defined by a hypersurface in a toric variety. It is shown  that these correspond one-to-one to different contributions to the line-bundle cohomology classes, i.e.~in particular the higher classes $H^i(X,L)$, $i>0$ can be identified with so-called twisted or non-geometric contributions. We show that an analogous correspondence also appears for the complete intersection of two hypersurfaces in a toric five-fold.

\vspace{0.5cm}
The high-performance {C/\CPP} implementation {\cohomCalg} of the algorithm \cite{cohomCalg:Implementation} including the recently added Koszul module is available under
\[
  \text{\href{http://wwwth.mppmu.mpg.de/members/blumenha/cohomcalg/}{http://wwwth.mppmu.mpg.de/members/blumenha/cohomcalg/}.}
\]

\section{Preliminaries}\label{sec_preliminaries}
This section begins with a summary of the algorithm for the computation of sheaf cohomology group dimensions in the setting of (complex) line bundles on toric varieties. Specific focus is given to the explicit mappings used in the determination of the multiplicity factors $\fh_i(\cQ)$.

Furthermore, before we can begin our survey of physically motivated applications, we need to define a basic ingredient: the tangent bundle for toric spaces. For a general manifold the tangent bundle is usually difficult to describe in terms which are usable for actual computations. However, for the case of toric varieties the situation improves a lot, see \cite{Cox:ToricVarieties, Fulton:ToricVarieties, Denef:2008wq, Kreuzer:2006ax, Reffert:2007im, HoriEA:MirrorSymmetry} for introductions to the subject. Here the tangent bundle can be described in terms of a short exact sequence, where the other two bundles are given by sums of line bundles. This is more generally known as the monad construction of vector bundles.

Via dualization we obtain the bundle of 1-forms in the same fashion and having such a description we can consider the exterior powers thereof, i.e.~the $n$-form bundles. Together with the holomorphic line bundle $\cO$ this allows us to compute the Hodge numbers of lower-dimensional toric spaces. In fact, those techniques can be extended to higher dimensions, but for the sake of a concise presentation we limit ourselves to cases of dimension up to five --- which is sufficient for nearly all applications in string model building.

A general technique that we are going to apply throughout this work is to consider the long exact cohomology sequence induced by a short exact sequence of bundles (or sheaves), see~\cite{Bredon:SheafTheory, Griffiths1994:Principles}. Provided that a sufficient number of the involved cohomology groups actually vanishes, i.e.~isomorphic to the trivial group, one can avoid working out the precise mappings of the sequences and instead argue directly on the basis of exactness. For actual computations this saves one extremely laborious computational step, as one only needs to consider the dimensions of the cohomology groups. We will demonstrate this in several explicit examples.

Furthermore, we acknowledge that a significant portion of the tools presented are not new and are scattered throughout the mathematics and physics literature, but in our opinion it might be useful for the reader to see all these methods collected on a couple of pages. We assume familiarity with the basic notions of toric geometry, sheaf and \v{C}ech cohomology to the extend of the material presented in our prior ``conjecture`` paper \cite{Blumenhagen:2010pv} \S2.1 as well as appendix A therein.

\subsection{Computing sheaf cohomologies of line bundles}\label{sec_basicalgorithm}
The computational algorithm used in this paper was first conjectured in \cite{Blumenhagen:2010pv} and subsequently proven in \cite{CohomOfLineBundles:Proof, Jow:2010arXiv1006.0780J}. The basic idea is to count certain rationoms, i.e.~rational functions with monomials both in the numerator and denominator, obtained from unions of the Stanley-Reisner ideal generators. Those individual contributions to the dimension of a sheaf cohomology group also have to be weighted differently, which basically catches the information of how a certain union monomial did arise. 

The generic setting throughout the paper is in the context of toric geometry. Let $X$ be a toric variety with the homogeneous coordinates $H=\{u_1,\dots,u_n\}$ and ${\rm SR}(X)$ the Stanley-Reisner ideal. We also require the gauged linear $\sigma$-model charges for each coordinate, which encode the projective relations. The Stanley-Reisner ideal consists of all square-free monomials whose coordinates are not contained in any cone of the toric fan $\Sigma_X$ and is Alexander-dual to the irrelevant ideal $B_\Sigma$, which is often used in the mathematical literature. 

Using this input data, the formula for the dimension of a sheaf cohomology group for a line bundle on a toric variety is
\beq\label{vflwolfburg}
  \boxed{ \dim H^i(X;\cO_X(D)) = \sum_{\cQ} \overbrace{\fh_i(\cQ)}^{\mclap{\text{multiplicity factor}}} \cdot \underbrace{\cN_D(\cQ)}_{\mclap{\text{number of rationoms}{}}}\qquad }
\eeq
where the sum ranges over all the square-free monomials $\cQ$ that can be obtained from unions of the Stanley-Reisner ideal generators.

\subsubsection*{Multiplicity factors}
The multiplicity factors $\fh_i(\cQ)$ themselves arise as the dimensions of an intermediate (relative) homology structure. For each subset $S^k_\alpha = \{ \cS_{\alpha_1},\dots,\cS_{\alpha_k}\} \subset{\rm SR}(X)$ let $\cQ(S^k_\alpha)$ denote the square-free monomial that arises from the union of all coordinates of all generators in this subset. Then we define
\beq
  N(S^k_\alpha) \ce | \cQ(S^k_\alpha) | - k,
\eeq
which measures sort of the multiplicity of coordinates appearing in more than one generator in a given subset of the Stanley-Reisner ideal. Using this number, we can define intermediate sequences $\fC_\bullet(\cQ)$ where the spaces are of dimension
\beq
  \dim \fC_i(\cQ) = \# \left\{ S^k_\alpha \subset {\rm SR} : \parbox{2.2cm}{$\cQ(S^k_\alpha) = \cQ$ \\ $N(S^k_\alpha) = i$} \right\},
\eeq
i.e.~the number of combinations of Stanley-Reisner ideal generators leading to the same square-free monomial $\cQ$ and $N(S^k_\alpha)$-value $i$.

We want to give an explicit description of the vector space structure of the $\fC_{i}(\cQ)$ and the maps between those, for details see Section 3 of \cite{CohomOfLineBundles:Proof}. Assuming that the Stanley-Reisner ideal is generated by $t$ different monomials as 
\beq
  {\rm SR}(X)=\langle \cS_{1},\dots ,\cS_{t}\rangle
\eeq
and setting $[t]\ce \lbrace 1, \dots ,t\rbrace$, we can define a relative complex $\Gamma^{\cQ}$ of the full simplex on $[t]$ by extracting only those subsets $\alpha\subset [t]$ with $\cQ(S_{\alpha}^{k})=\cQ$. For some fixed cardinality $\vert \alpha \vert = k $, we define the set of $(k-1)$-dimensional\footnote{Note that a subset with a single element is a face of dimension $0$ and the empty set $\emptyset\subset [t]$ is a face of dimension $-1$ in this formalism.} faces $F_{k-1}(\cQ)$ of $\Gamma^{\cQ}$ and let $\IC^{F_{k-1}(\cQ)}$ be the complex vector space with basis vectors $e_{\alpha}$ corresponding to $\alpha\in F_{k-1}(\cQ)$. The (relative) complex 
\beq
  F_{\bullet}(\cQ):\quad 0 \fto F_{t-1}(\cQ) \stackrel{\phi_{t-1}}{\fto} \cdots
  \stackrel{\phi_1}{\fto} F_0 (\cQ ) \stackrel{\phi_{0}}{\fto} F_{-1}(\cQ) \fto 0
\eeq
is then given by the maps 
\beq 
  \phi_k: F_k(\cQ) \fto F_{k-1}(\cQ) \, ,\quad e_{\alpha} \mapsto \sum_{s\in \alpha} \text{sign}(s,\alpha) e_{\alpha\setminus s}\, , 
\eeq 
where $e_{\alpha\setminus s}=0$ if $\alpha\setminus s\notin \Gamma^{\cQ}$ and $\text{sign}(s,\alpha )=(-1)^{\ell -1 }$ when $s$ is the $\ell$-th element of $\alpha\subset [t]$ written in increasing order. The complex $\fC_{\bullet}(\cQ)$ is just a relabeling of $F_{\bullet}(\cQ)$, i.e.~one sets $\fC_{i}(\cQ)=F_{\vert \cQ \vert - i}(\cQ)$ while leaving the maps untouched.

The multiplicity factors then correspond to the homology dimensions of the complex $\fC_\bullet(\cQ)$. Note that the $\fh_i(\cQ)$ only depend on the geometry of the toric variety $X$, but not on the divisor $D$ that specifies the line bundle $\cO_X(D)$. It therefore suffices to compute the multiplicities $\fh_i(\cQ)$ only once for any given geometry.

\subsubsection*{Counting rational functions}
The second part of the algorithm depends on the GLSM charge of the divisor $D$ that determines the line bundle $\cO_X(D)$. Given a square-free monomial $\cQ=x_{i_1}\cdots x_{i_k}$, where $I=(i_1,\dots,i_k,\dots,i_n)$ is an index renumbering such that the product of the first $k$ coordinates gives the monomial $\cQ$, we consider rational functions of the form
\beq\label{eq_rationomprototype}
  R^\cQ(x_1,\dots,x_n) = \frac{T(x_{k+1},\dots,x_{n})}{x_{i_1}\cdots x_{i_k} \cdot W(x_{i_1},\dots,x_{i_k})},
\eeq
where $T$ and $W$ are monomials. Therefore, the coordinates in $\cQ$ appear in the denominator, whereas the remaining coordinates (the complement of $\cQ$ in $H$) are found in the numerator. If we take the GLSM charges of the numerator positive and of the denominator negative we can define
\beq
  \cN_D(\cQ) \ce \dim \{ R^\cQ : \deg R^\cQ = D \},
\eeq
which effectively counts the number of rational functions whose GLSM degree is equal to the divisor GLSM charges defining the line bundle. Together with the multiplicity factors, the sum of all such contributions gives the dimension of the sheaf cohomology groups for $\cO_X(D)$.

It should be mentioned that a different algorithm for the same problem has been known for some time, see section~9.1 of \cite{Cox:ToricVarieties}. It has been implemented and applied in \cite{Cvetic:2010rq}. However, due to the entirely different methods of the computation (subdividing and scanning lattices versus computing unions of Stanley-Reisner ideal generators and counting rationoms) this algorithm quickly becomes computationally expensive.

\subsection{The tangent bundle}
Another crucial ingredient is the tangent bundle, which for toric varieties can be described via the general monad bundle construction (see next section for the general case). Let $D_k\in{\rm Div}(X)$ be the vanishing divisor $\{ u_k = 0 \}$ associated to each homogeneous coordinate. Then there is the generalized short exact Euler sequence
\beq\label{eq_generalizedeuler}
  0 \fto \cO_X^{\oplus r} \injto \bigoplus_{i=1}^n \cO_X(D_k) \surjto T_X \fto 0,
\eeq
which relates the holomorphic line bundles on $X$ to the tangent bundle, where $r$ is the number of projective relations defining the toric variety. From this we are considering the induced long exact sequence of sheaf cohomology groups
\beq\label{eq_longexacttangent}
  \parbox{1cm}{\xymatrix{ 
      0 \ar[r] & H^0(X;\cO_X)^{\oplus r} \ar[r] & \smash{\bigoplus\limits_{k=1}^n H^0(X;\cO_X(D_k))} \ar[r] & H^0(X;T_X) \ar`[rd]`[l]`[dlll]`[d][dll] & \\
               & H^1(X;\cO_X)^{\oplus r} \ar[r] & \smash{\bigoplus\limits_{k=1}^n H^1(X;\cO_X(D_k))} \ar[r] & H^1(X;T_X) \ar`[rd]`[l]`[dlll]`[d][dll] & \\
               & H^2(X;\cO_X)^{\oplus r} \ar[r] & \smash{\bigoplus\limits_{k=1}^n H^2(X;\cO_X(D_k))} \ar[r] & H^2(X;T_X) \ar[r]                       & \dots}}
\eeq
and via computing $H^i(X;\cO_X)$ as well as $H^i(X;\cO_X(D_k))$ we may derive $H^i(X;T_X)$ from exactness provided that enough cohomology groups are vanishing.

In order to compute the Hodge diamond of some complex surface or threefold using \eqref{eq_hodgeomegaiso}, one dualizes the sequence \eqref{eq_generalizedeuler}. This gives
\beq\label{eq_dualgeneralizedeuler}
  \boxed{0 \fto T_X^* \cong \Omega^1_X \injto \bigoplus_{k=1}^n\cO_X(-D_k) \surjto \cO_X^{\oplus r} \fto 0,}
\eeq
which is again a short exact sequence, so one may consider the induced long exact sequence in order to derive $H^{1,i}(X)\cong H^i(X;\Omega^1_X)$ from $H^i(X;\cO_X)$ and $H^i(X;\cO_X(D_k))$ --- this suffices to derive the entire Hodge diamond $H^{p,q}(X)$ for $\dim_\IC X \le 3$ using the usual symmetries of the Hodge diamond.

\subsubsection*{Example: Hodge diamond of Hirzebruch surfaces}
\begin{table}[t]
  \centering
  \begin{tabular}{r@{\,$=$\,(\,}r@{,\;\;}r@{\,)\;\;}|c|cc|c}
    \multicolumn{3}{c|}{vertices of the} & coords & \multicolumn{2}{c|}{GLSM charges} & {divisor class}${}^\big.$ \\
    \multicolumn{3}{c|}{polyhedron / fan} & & $Q^1$ & $Q^2$ & \\
    \hline\hline
    $\nu_1$ & $-1$ & $-k$ & $u_1$ & 1   & 0 & $H$\uspc \\
    $\nu_2$ &  1   &  0   & $u_2$ & 1   & 0 & $H$  \\
    $\nu_3$ &  0   &  1   & $u_3$ & $k$ & 1 & $kH+X$ \\
    $\nu_4$ &  0   & $-1$ & $u_4$ & 0   & 1 & $X$
  \end{tabular}
  \\[2mm] $\text{intersection form:}\quad HX -k X^2$
  \\[2mm] ${\rm SR}(F_k) = \langle u_1 u_2 ,\; u_3 u_4 \rangle$
  \caption{\small Toric data for the Hirzebruch surface $F_k$}
  \label{tab_HirzebruchSurface}
\end{table}

Let us apply this method to the Hirzebruch surfaces $F_k$, which are $\IP^1$-bundles over $\IP^1$ twisted according to the sheaf $\cO(0)+\cO(-k)$. This in particular includes the well-known del Pezzo special cases $F_0 = \IP^1\times\IP^1$ and $F_1 = dP_1$, which is $\IP^2$ blown up at a single point. For some divisor $D=nH+mX$ we use the notation $\cO_{F_k}(D)=\cO(n,m)$. The relevant toric information for the Hirzebruch surfaces $F_k$ is provided in Table \ref{tab_HirzebruchSurface}. The dualized generalized Euler sequence \eqref{eq_dualgeneralizedeuler} then reads
\beq\label{eq_hirzebruchseq}
  0 \fto \Omega^1_{F_k} \injto \cO(-1,0)^{\oplus 2}\oplus\cO(-k,-1)\oplus\cO(0,-1) \surjto \cO(0,0)^{\oplus 2} \fto 0.
\eeq
After one makes the effort to compute the cohomology $H^\bullet(F_k;\cO(m,n))$ for the four required values $(m,n)=(0,0)$, $(-1,0)$, $(-k,-1)$ and $(0,-1)$, which gives
\beq\label{eq_hirzebruchintcohoms}
  \bal
    & h^\bullet(F_k;\cO(0,0))  \cong h^{0,\bullet}(F_k) = (1,0,0) \\
    & h^\bullet(F_k;\cO(-1,0)) = h^\bullet(F_k;\cO(-k,-1)) = h^\bullet(F_k;\cO(0,-1)) = (0,0,0)     
  \eal
\eeq
and therefore $h^\bullet(F_k; \cO(-1,0)^{\oplus 2}\oplus\cO(-k,-1)\oplus\cO(0,-1)) = (0,0,0)$, the induced long exact sheaf cohomology sequence of \eqref{eq_hirzebruchseq} is considered. Note that in general $H^i(X;\Omega^1_X) \cong \IR^{h^i(X;\Omega^1_X)}$, which gives
\beq
  \parbox{1cm}{\xymatrix{ 
    0 \ar[r] & H^0(F_k;\Omega^1_{F_k}) \ar[r] & \IR^0=0 \ar[r] & \IR^{1\cdot 2}=\IR^2 \ar`[rd]`[l]`[dlll]_{\cong}`[d][dll] & \\
             & H^1(F_k;\Omega^1_{F_k}) \ar[r] & \IR^0=0 \ar[r] & \IR^0=0 \ar`[rd]`[l]`[dlll]`[d][dll] & \\
             & H^2(F_k;\Omega^1_{F_k}) \ar[r] & \IR^0=0 \ar[r] & \IR^0=0 \ar[r]                       & 0.}}
\eeq
The sequence effectively terminates due to dimensional reasons, i.e.~all remaining cohomology groups are trivial. From the exactness of the sequence we may therefore deduce
\beq\label{eq_hirzebruchHone}
  h^\bullet(F_k;\Omega^1_{F_k}) \cong h^{1,\bullet}(F_k) = (0,2,0).
\eeq
This completes the computation and via the usual dualities and relations between the Hodge numbers (basically, it suffices to determine just one quadrant of the Hodge diamond) we obtain from \eqref{eq_hirzebruchintcohoms} and \eqref{eq_hirzebruchHone} the Hirzebruch surface's Hodge diamond
\beq\label{eq_hirzebruchseqex}
  \bal
    \begin{array}{ccccc}
                &                 & h^{0,0} \\
                & \mclap{h^{1,0}} &         & \mclap{h^{0,1}} \\
        h^{2,0} &                 & h^{1,1} &         & h^{0,2} \\
                & \mclap{h^{1,0}} &         & \mclap{h^{0,1}} \\
                &                 & h^{0,0}
      \end{array}
      \quad&=\quad
    \begin{array}{ccccc}
                &                 & h^0(\cO_X) \\
                & \mclap{h^0(\Omega^1_X)} &         & \mclap{h^1(\cO_X)} \\
        h^{2,0} &                 & h^1(\Omega^1_X) &         & h^2(\cO_X) \\
                & \mclap{h^{1,0}} &         & \mclap{h^2(\Omega^1_X)} \\
                &                 & h^{0,0}
      \end{array} \\
      \quad&=\quad
    \begin{array}{ccccc|c}
        &   & 1 &   &   & b^0=1 \\
        & 0 &   & 0 &   & b^1=0 \\
      0 &   & 2 &   & 0 & b^2=2 \\
        & 0 &   & 0 &   & b^3=0 \\
        &   & 1 &   &   & b^4=1
    \end{array}
    \eal
\eeq
in perfect agreement with the expected results. The dimensions of the cohomology groups $H^i(F_k;\cO(m,n))$ in \eqref{eq_hirzebruchintcohoms} can be quickly computed using the implementation \cite{cohomCalg:Implementation} of the algorithm.

\subsection{The monad and extension bundle construction}
The Euler sequence \eqref{eq_generalizedeuler} and its dual \eqref{eq_dualgeneralizedeuler} are specific examples  of a more general construction known as monad bundles. Here a bundle is indirectly defined via a short exact sequence with  two other known bundles. Those are usually Whitney sums of line bundles for computational simplicity. The general structure is therefore
\beq\label{eq_monadbundleprototype}
  0 \fto V \stackrel{f}{\injto} \bigoplus_{i=1}^{r_B} \cO_X(b_i) \stackrel{g}{\surjto} \bigoplus_{i=1}^{r_C} \cO_X(c_i) \fto 0
\eeq
which via dualization and changing the signs of the individual line bundles also implies
\beq\label{eq_dualmonadbundleprototype}
  0 \fto \bigoplus_{i=1}^{r_A} \cO_X(a_i) \injto \bigoplus_{i=1}^{r_B} \cO_X(b_i) \surjto U \fto 0 .
\eeq
The exactness of both sequences necessarily implies the bundle isomorphisms
\beq
  \bal
	  & U \cong \bigoplus_{i=1}^{r_B} \cO_X(b_i) \bigg/ \bigoplus_{i=1}^{r_A}\cO_X(a_i), \\
		& \bigoplus_{i=1}^{r_B} \cO_X(b_i) \bigg/ V \cong \bigoplus_{i=1}^{r_C}\cO_X(c_i),
	\eal
\eeq
which shows that monad bundles are closely related to coset space constructions at bundle niveau. 

The idea has been used widely in the construction of suitable (gauge) vector bundles for heterotic compactification and directly appears in the generalization of the two-dimensional gauged linear sigma model to the  $(0,2)$  supersymmetric case\cite{Witten:1993yc}. Indeed, one does allow the left-moving world-sheet fermions to couple not to the tangent bundle of the CY manifold (standard embedding) but to a more general (stable) holomorphic vector bundle of in general  rk$(V)\ne 3$. This allows for more general observable gauge groups $G\ne E_6$ like the GUT groups $SO(10)$ or $SU(5)$.  

At this point we simply would like to highlight the importance of line bundle cohomologies for the computation of cohomologies involving monad bundles. From the long exact cohomology sequence that is induced from the monad sequences one can easily determine the monad bundle cohomology from the known line bundle cohomologies.

A closely related approach is the so-called extension bundle construction, where the indirectly defined bundle sits in the middle of the short exact sequence
\beq
  0 \fto V_1 \stackrel{\tilde f}{\injto} W \stackrel{\tilde g}{\surjto} V_2 \fto 0\; ,
\eeq
where for getting a non-split extension one also requires $H^1(X, V_1\otimes V^*_2)\ne 0$. Again, often the two bundles $V_1$ and $V_2$ are chosen to be direct sums of line bundles
\beq\label{eq_extensionsbundleprototype}
  0 \fto \bigoplus_{i=1}^{r_A} \cO_X(a_i) \stackrel{\tilde f}{\injto} W \stackrel{\tilde g}{\surjto} \bigoplus_{i=1}^{r_C} \cO_X(c_i) \fto 0\; .
\eeq
Clearly, the rank of the monad and the extension bundles are given by
\beq
  \bal
    &\text{monad bundles: } &&\bal {\rm rk}(V) &= r_B - r_C \\ {\rm rk}(U) &= r_B - r_A \eal \\
	  &\text{extension bundles: } && {}\,{\rm rk}(W) = r_A + r_B\;  .
	\eal
\eeq

\subsection{Hodge numbers of 4- and 5-dimensional toric spaces}\label{sec_highdimambienthodgenumbers}
In order to compute the Hodge diamond for higher-dimensional spaces, we need higher exterior powers of the cotangent sheaf, i.e.~we require $\Omega^p_X$ for $p>1$. We restrict to the case of 4- and 5-folds, such that only the sheaf / bundle of 2-forms $\Omega^2_X$ is actually required. In this case, there are a number of general sequences, which allow to derive the antisymmetric tensor product. Let 
\beq
  0\fto A \injto B \surjto C \fto 0
\eeq
be a short exact sequence of vector bundles or sheaves. Then all four of the following sequences are short and exact as well:
\beq
  \parbox{1cm}{\xymatrix{ & & 0 \ar[d] & 0 \ar[d] & \\
                        0 \ar[r] & \Lambda^2A \ar@{^{(}->}[r] & Q_1 \ar@{->>}[r] \ar@{^{(}->}[d] & A\otimes C \ar[r] \ar@{^{(}->}[d] & 0 \\
                        0 \ar[r] & \Lambda^2A \ar@{^{(}->}[r] & \Lambda^2 B \ar@{->>}[d] \ar@{->>}[r] & Q_2 \ar[r] \ar@{->>}[d] & 0 \\
                        & & \Lambda^2 C \ar[d] & \Lambda^2 C \ar[d] & \\
                        & & 0 & 0 & }}.
\eeq
This basically yields two ways to compute $\Lambda^2A$, $\Lambda^2B$ and $\Lambda^2C$ using the two split short exact sequences
\beq\label{eq_lambdatwovarA}
  \bal
    & 0 \fto \Lambda^2A \injto Q_1 \surjto A\otimes C \fto 0 \\
    & 0 \fto Q_1 \injto \Lambda^2B \surjto \Lambda^2C \fto 0
  \eal
\eeq
or the second pair
\beq\label{eq_lambdatwovarB}
  \bal
    & 0 \fto \Lambda^2A \injto \Lambda^2B \surjto Q_2 \fto 0 \\
    & 0 \fto A\otimes C \injto Q_2 \surjto \Lambda^2C \fto 0.
  \eal
\eeq
Choosing the right pair depends on the ability to make use of exactness. If either $A$ or $C$ is a rank-1 bundle or sheaf --- which implies either $\Lambda^2A=0$ or $\Lambda^2C=0$, such that the remaining part of the sequence provides an actual isomorphism for the auxiliary bundle $Q_i$ --- only a single short exact sequence remains:
\beq\label{eq_lambdatwolinebdls}
  \bal
    \text{$A$ line bundle:} &{\qquad}& 0 \fto A\otimes C \injto \Lambda^2B \surjto \Lambda^2C \fto 0 \\
    \text{$C$ line bundle:} && 0 \fto \Lambda^2A \injto \Lambda^2B \surjto A\otimes C \fto 0
  \eal
\eeq

This general approach is now applied to the dualized general Euler sequence \eqref{eq_dualgeneralizedeuler}, such that $A=\Omega^1_X$, $B=\bigoplus_{k=1}^n \cO_X(-D_k)$ and $C=\cO_X^{\oplus r}$ are used. First, the second short exact sequence of \eqref{eq_lambdatwovarB} becomes
\beq\label{eq_lambdatwoambient1}
  0 \fto (\Omega^1_X)^{\oplus r} \injto Q_2 \surjto \cO_X^{\oplus\binom{r}{2}} \fto 0,
\eeq
which allows to derive the cohomology $H^\bullet(X;Q_2)$ of the auxiliary sheaf / bundle $Q_2$ via the usual method of using exactness of the induced long exact sequence. This is then used in the first sequence of \eqref{eq_lambdatwovarB}, i.e.~
\beq\label{eq_lambdatwoambient2}
  0 \fto \Omega^2_X \injto \bigoplus_{i<j} \cO_X(-D_i-D_j) \surjto Q_2 \fto 0,
\eeq
to obtain $H^\bullet(X;\Omega^2_X)$. Naturally, one may use the other set of sequences \eqref{eq_lambdatwovarA} as an alternative way, which requires running through the sequences
\beq
  \bal
    & 0 \fto Q_1 \injto \bigoplus_{i<j} \cO_X(-D_i - D_j) \surjto \cO_X^{\oplus\binom{r}{2}} \fto 0 \\
    & 0 \fto \Omega^2_X \injto Q_1 \surjto (\Omega^1_X)^{\oplus r} \fto 0.
  \eal
\eeq
Choosing the right set of sequences depends mostly on the ability to extract information purely from the exactness of the induced long exact cohomology sequences, which crucially relies on the appearance of ideally lots of zeros in the known cohomologies.

It should be noted that the previous two short exact sequences simplify if we are restricting to the case of a weighted projective space with just a single projection relation (i.e.~$r=1$) between the coordinates. In that case the first sequence \eqref{eq_lambdatwoambient1} establishes $Q_2\cong\Omega^1_X$, such that we obtain the short exact sequence
\beq
  0 \fto \Omega^2_X \injto \bigoplus_{i<j} \cO_X(-D_i-D_j) \surjto \Omega^1_X \fto 0
\eeq
from \eqref{eq_lambdatwoambient2}, which can be used in the same fashion as before in order to derive the cohomology of the sheaf $\Omega^2_X$. This result also agrees with the second sequence of \eqref{eq_lambdatwolinebdls}.

Using the general symmetries and dualities between the Hodge numbers, the cohomology of $\cO_X$, $\Omega^1_X$ and $\Omega^2_X$ suffices to determine the entire Hodge diamond of up to 5-folds. Deriving even higher exterior powers of the cotangent bundle $\Omega^p_X$ for $p\ge 3$ is in principle similar, but becomes way more complicated due to further splittings of the underlying long exact sequence of the exterior powers. Due to a lack of actual applications, we will not dwell on this issue any further.

\section[Cohomology for Calabi-Yau manifolds and D-branes]{Cohomology for Calabi-Yau manifolds \\ and D-branes}\label{sec_calabiyau}
The ability to algorithmically compute the line bundle sheaf cohomologies of toric spaces covers a great variety of geometries, e.g.~weighted projective spaces, the lower-degree del Pezzo surfaces $dP_1$, $dP_2$ and $dP_3$, the Hirzebruch surfaces and many others. However, for applications to supersymmetric compactifications of the ten-dimensional superstring theories it is necessary to work with compact Calabi-Yau manifolds. Such compact spaces have vanishing first Chern class and therefore  are not simply given by toric varieties. However, hypersurfaces and more general complete intersections of hypersurfaces in toric varieties can define genuine Calabi-Yau manifolds and in fact constitute the largest known class of such spaces. The classic example is a degree five hypersurface in $\IP^4$, also known as the quintic.

For the heterotic string, in addition to the CY manifold $X$ one also needs to specify a stable holomorphic vector bundle $V$ satisfying the tadpole cancellation condition 
\beq\label{tadpole}
  {\rm ch}_2(V)+c_2(T_X)=\sum_a N_a \gamma_a\, ,
\eeq
where $N_a$ denotes the number of five-branes wrapping a holomorphic, effective two-cycle Poincar\'e dual to the closed four-form $\gamma_a$. The structure group $G$ of the vector bundle is embedded into the $E_8\times E_8$ respectively $SO(32)$ gauge group of the ten-dimensional heterotic string, breaking it to a smaller observable gauge group $H$. In this respect the $E_8\times E_8$ heterotic string is considered very encouraging, as for the structure groups $G=SU(3)$, $SU(4)$, $SU(5)$ one gets the observable ones $H=E_6$, $SO(10)$, $SU(5)$. These are all candidate GUT groups. In addition one generically gets matter fields, which, depending on the decomposition $G\times H\subset E_8$, transform in various representations of the observable gauge group $H$. Their number is determined by the cohomology groups $H^i(X,\wedge^n V)$, $i=1,2$. In addition there are gauge singlets from the vector bundle deformations counted by $H^1(X, {\rm End}(V))$. A simple solution to the tadpole condition \eqref{tadpole} is to choose $V=T_X$ and $N_a=0$, which leads to $H=E_6$ and matter fields in the ${\bf 27}/{\bf\overline{27}}$ representation counted by $H^1(X,T_X)=H^{2,1}(X)$ respectively $H^2(X,T_X)=H^{1,1}(X)$. In this section, we will mainly consider this latter case, but it will be clear that our methods straightforwardly apply also to more general vector bundles $V$.

The other large class of string compactifications are orientifolds of the Type IIA/B superstring. Here one also compactifies the ten-dimensional string on a Calabi-Yau manifold and then takes a quotient $\Omega\sigma$, where $\Omega$ denotes the world-sheet parity transformation and $\sigma$ a geometric $\IZ_2$ involution of $X$. In general this leads to lower-dimensional orientifold planes, whose tadpole needs to be canceled by corresponding lower-dimensional D-branes wrapping certain cycles of $X$. In this case the gauge group is supported on the D-branes and the (chiral) matter on mutual intersections of two D-branes. Thus, here one also needs to compute cohomology classes on certain subspaces of the Calabi-Yau manifold. As long as these subspaces are divisors or complete intersections of them our methods also apply.

So far we have been able to efficiently compute the sheaf cohomology of line bundles defined over the toric ambient space itself. The mathematical interconnection to subspaces of the aforementioned type is the Koszul complex, which in its most basic formulation directly relates the ambient space cohomology to the hypersurface cohomology. Provided that further restrictions on the geometry are mutually compatible (``transverse``) and leading to a well-defined subspace, the case of complete intersections can be handled by repeated application of this sequence. In the end, this provides a fully algorithmic method to determine the line bundle sheaf cohomology of toric subspaces.

\subsection{The Koszul sequence}\label{sec:koszulsequence}
Consider an irreducible hypersurface $D\subset X$ and let $0\not=\sigma\in H^0(X;\cO(D))$ be a global nonzero section of $\cO_X(D)$ such that $Z(\sigma)\cong D$. This induces a mapping $\cO_X\fto\cO_X(D)$ and its dual $\cO_X(-D)\fto\cO_X$, the latter of which can be shown to be injective.

This statement can be extended by computing the image of the mapping. Given any effective divisor $D=\sum_i a_i[H_i]\in{\rm Div}(X)$, i.e.~all $a_i\ge0$, there is a short exact sequence
\beq\label{eq_koszulsequence}
  \boxed{0 \fto \cO_X(-D) \injto \cO_X \surjto \cO_D \fto 0},
\eeq
sometimes called the Koszul sequence, where $\cO_D$ is the quotient of the sheaf $\cO_X$ of holomorphic functions on $X$ by all holomorphic functions vanishing at least to order $a_i$ on the irreducible hypersurface $H_i$. In particular, $\cO_D$ can be regarded as the structure sheaf on $D\subset X$, which effective yields $H^i(X;\cO_D)\cong H^i(D;\cO_D)$. The Koszul sequence is of utmost importance in the subsequent sections, as its induced long exact sheaf cohomology sequence
\beq\label{eq_longexactkoszul}
  \parbox{1cm}{\xymatrix{ 
      0 \ar[r] & H^0(X;\cO_X(-D)) \ar[r] & H^0(X;\cO_X) \ar[r] & H^0(D;\cO_D) \ar`[rd]`[l]`[dlll]`[d][dll] & \\
               & H^1(X;\cO_X(-D)) \ar[r] & H^1(X;\cO_X) \ar[r] & H^1(D;\cO_D) \ar`[rd]`[l]`[dlll]`[d][dll] & \\
               & H^2(X;\cO_X(-D)) \ar[r] & H^2(X;\cO_X) \ar[r] & H^2(D;\cO_D) \ar[r]                       & \dots}}
\eeq
allows to relate the ambient space cohomology (left and middle column) to the cohomology of the divisor (right column). The practical usage of this sequence requires most of the cohomology classes to vanish, such that one can use the exactness to deduce isomorphisms between the cohomology groups or their vanishing without having to bother about the mappings.\footnote{See \cite{Griffiths1994:Principles} for a full mathematical account on the Koszul complex, in particular the mappings.}

However, the plain Koszul sequence and \eqref{eq_longexactkoszul} only allow to compute the holomorphic cohomology of the divisor $D$. Fortunately the $\cO_X(T)$-twisted (i.e.~tensored) version of the Koszul sequence
\beq
  0 \fto \cO_X(-D+T) \injto \cO_X(T) \surjto \cO_D(T) \fto 0
\eeq
is exact as well, which allows to compute the line bundle cohomology on divisors as well. Aside from that recall the isomorphism
\beq\label{eq_hodgeomegaiso}
  H^{p,q}(X) \cong H^q(X;\Omega^p_X),
\eeq
which relates the Dolbeault cohomology groups $H^{p,q}$ to the (sheaf) cohomology with values in the bundle $\Omega^p_X$ of $(p,0)$-forms on $X$. In particular, since $\Omega^0_X \cong \cO_X$ it immediately follows
\beq
  H^i(X;\cO_X) \cong H^{0,i}(X),
\eeq
so we are actually computing the ``edge'' of the Hodge diamond of $X$, which is of somewhat limited value for the entire topology.

\subsection{Hypersurfaces}\label{subsection:Hypersurface}
At this point we already know how to construct the tangent bundle of the ambient toric space. Unfortunately, the situation becomes much more involved if we are interested in the tangent bundle of a hypersurface inside some toric variety. Let $D$ denote the divisor class of the hypersurface and $D_k$ as before. Then the tangent bundle can be derived as the measure of non-exactness of the sequence
\beq\label{eq_tangentbundlehypersurface}
  0 \fto \cO_D^{\oplus r} \stackrel{\alpha}{\injto} \bigoplus_{k=1}^n \cO_D(D_k) \stackrel{\beta}{\fto} \cO_D(D),
\eeq
i.e.~with respect to the mappings $\alpha$ and $\beta$ we have
\beq
  T_{D} = \frac{{\rm ker}(\beta)}{{\rm Image}(\alpha)},
\eeq
which defines a quotient bundle of $\bigoplus_{k=1}^n\cO_D(D_k)$. This particular definition --- albeit formally elegant --- does not help if one tries to actually compute the tangent bundle cohomology. Therefore we employ an auxiliary sheaf $\cE_D$ on the hypersurface $D$ which is defined indirectly such that the sequences
\beq\label{eq_splittangentbundle}
  \bal
    &0 \fto \cO_D^{\oplus r} \injto \bigoplus_{k=1}^n \cO_D(D_k) \surjto \cE_D \fto 0 \\
    &0 \fto T_D \injto \cE_D \surjto \cO_D(D) \fto 0
  \eal
\eeq
are exact, i.e.~we effectively represent the definition of the hypersurface's tangent bundle by a split into two exact sequences. From another perspective the sheaf $\cE_D$ can be identified with the restriction of the tangent bundle sheaf of the ambient space $X$, i.e.~one may treat $\cE_D$ like $T_X|_D$. 

Thus, following the by now established method of using the exactness of the induced long exact cohomology sequences we first may determine the sheaf cohomology $H^i(D;\cE_D)$ from the first sequence and then run through the second sequence to determine $H^i(D;T_D)$. It is also necessary to compute the restrictions of $\cO_X(D_k)$ to $D$, which is accomplished via tensoring the Koszul sequence \eqref{eq_koszulsequence} with the divisor $\cO_X(D_k)$, i.e.
\beq\label{eq_koszultensored}
  0 \fto \underbrace{\cO_X(D_k)\otimes\cO_X(-D)}_{\cO_X(D_k-D)} \injto \underbrace{\cO_X(D_k)\otimes\cO_X}_{\cO_X(D_k)} \surjto \underbrace{\cO_X(D_k)\otimes\cO_D}_{\cO_D(D_k)} \fto 0.
\eeq
Obviously, those computations become quite expensive for higher-dimensional varieties or sufficiently complex toric ambient spaces, as we have to run through several long exact sequences.

All the short exact sequences can be dualized in order to derive the cohomology of the sheaf $\Omega^1_D$, which is necessary to compute the Hodge diamond of some surface or threefold. This yields the three sequences
\beq\label{eq_dualhypersurfacetangent}
  \bal
    & 0 \fto \cO_X(D_k-D) \injto \cO_X(D_k) \surjto \cO_D(D_k) \fto 0 \\
    & 0 \fto \cE_D^* \injto \bigoplus_{k=1}^n \cO_D(-D_k) \surjto \cO_D^{\oplus r} \fto 0 \\
    & 0 \fto \cO_D(-D) \injto \cE_D^* \surjto T_D^* \cong \Omega^1_D \fto 0
  \eal
\eeq
in order to determine the cohomology of the sheaves $\cO_D(D_k)$ to subsequently compute
the cohomology of $\cE_D^*$ and $\Omega^1_D$.

\subsubsection*{\texorpdfstring{Example: Hodge diamond of the octic $\IP^4_{11222}[8]$}{Example: Hodge diamond of the octic P11222[8]}}
To exemplify the discussed methods, we compute the Hodge diamond of the embedded Calabi-Yau hypersurface $\IP^4_{11222}[8]$. Since the weighted projective ambient space $\IP^4_{11222}$ has a $\IZ_{2}$-singularity and our methods only apply for smooth ambient spaces, we first torically blow it up to get the toric data of the smooth ambient space $X$ as shown in Table \ref{tbl:TorBlowP11222}.

\begin{table}[ht]
  \begin{center}
    \begin{tabular}{r@{\,$=$\,(\,}r@{,\;\;}r@{,\;\;}r@{,\;\;}r@{\,)\;\;}|c|cc|c} 
      \multicolumn{5}{c|}{vertices of the} & coords & \multicolumn{2}{c|}{GLSM charges} & {divisor class}${}^\big.$ \\
      \multicolumn{5}{c|}{polyhedron / fan}      &        & $Q^1$ & $Q^2$ & \\ 
      \hline\hline
      $\nu_1$ &   $-1$  &   $-2$  &   $-2$  &   $-2$  &    $u_1$   & 1 & 0  & $H$\uspc \\
      $\nu_2$ &    1    &    0    &    0    &    0    &    $u_2$   & 1 & 0  & $H$ \\
      $\nu_3$ &    0    &    1    &    0    &    0    &    $u_3$   & 2 & 1  & $2H+X$ \\
      $\nu_4$ &    0    &    0    &    1    &    0    &    $u_4$   & 2 & 1  & $2H+X$ \\
      $\nu_5$ &    0    &    0    &    0    &    1    &    $u_5$   & 2 & 1  & $2H+X$ \\
      $\nu_6$ &    0    &   $-1$  &   $-1$  &   $-1$  &    $u_6$   & 0 & 1  & $X$\lspc \\
      \hline
      \multicolumn{6}{c|}{conditions:}                          & 8 & 4  & \uspc  \\ 
    \end{tabular}
        \\[2mm] ${\rm SR}(\widetilde{\IP}^4_{11222}) = \langle u_{1} u_{2} ,\; u_3 u_4 u_5 u_{6} \rangle$
        \\[2mm] $\begin{aligned} \Sigma(\widetilde{\IP}^4_{11222}) = \big\langle & [2\,3\,4\,5],\; [1\,3\,4\,5],\; [2\,3\,4\,6],\; [2\,4\,5\,6],\; 
                                                                            \\ & [2\,3\,5\,6],\; [1\,3\,4\,6],\; [1\,4\,5\,6],\; [1\,3\,5\,6] \big\rangle \end{aligned}$
  \end{center}
  \caption{\small The torically blown-up weighted projective space $X\ce\widetilde{\IP}^4_{11222}$ with embedded CY-hypersurface given by the divisor $D=8H+4X$ with charges $(8,4)$.}
  \label{tbl:TorBlowP11222}
\end{table}

\vspace{0.3cm}
\noindent
We will look at the hypersurface in $X$ given by the divisor $D = (8,4)$ and compute its Hodge diamond. This will be the same as the Hodge diamond of $\IP^4_{11222}[8]$, since this hypersurface misses the $\IZ_{2}$-singularity anyway, i.e.~the blow-up is just necessary to make use of the algorithm.

Inserting the data into the last two sequences of \eqref{eq_dualhypersurfacetangent}, we get 
\beq\label{eq_quinticseqex} 
  \bal 
    &0 \fto \cE_D^* \injto \cO_D(-1,0)^{\oplus 2}\oplus \cO_D(-2,-1)^{\oplus 3}\oplus\cO_D(0,-1) \surjto \cO_D^{\oplus 2} \fto 0 \\
    &0 \fto \cO_D(-8,-4) \injto \cE_D^* \surjto T_D^* \cong \Omega^1_D \fto 0. 
  \eal 
\eeq 
To make use of these, it is necessary to determine the cohomology of $\cO_D(-1,0)$, $\cO_D(-2,-1)$, $\cO_D(0,-1)$, $\cO_D$ and $\cO_D(-8,-4)$, which is done using the first sequence in \eqref{eq_dualhypersurfacetangent}. For example, to get $\cO_D(-1,0)$, one takes the short exact sequence 
\beq 
  0 \fto \cO_{X}(-9,-4) \injto \cO_{X}(-1,0) \surjto \cO_D(-1,0) \fto 0 
\eeq 
and then looks at the long exact sequence in cohomology. Therefore, it is sufficient to know the cohomology of the ambient space line bundles $\cO_{X}(a)$ with divisor charges 
\beq 
  \bal 
    a\in\big\{ &(-9,-4),(-1,0),(-10,-5),(-2,-1),\\
		&(-8,-5),(0,-1),(-8,-4),(0,0),(-16,8)\big\}\, ,
  \eal 
\eeq
for which our algorithm yields the cohomology group dimensions
\beq 
  \bal 
    & h^\bullet(X;\cO_{X}(-9,-4)) {}= (0,0,0,0,2) , \quad &&{} h^\bullet(X;\cO_{X}(-1,0)) {}= (0,0,0,0,0) , \\
    & h^\bullet(X;\cO_{X}(-10,-5)){}= (0,0,0,0,6) , \quad &&{} h^\bullet(X;\cO_{X}(-2,-1)){}= (0,0,0,0,0) , \\
    & h^\bullet(X;\cO_{X}(-8,-5)) {}= (0,0,0,3,1) , \quad &&{} h^\bullet(X;\cO_{X}(0,-1)) {}= (0,0,0,0,0) , \\
    & h^\bullet(X;\cO_{X}(-8,-4)) {}= (0,0,0,0,1) , \quad &&{} h^\bullet(X;\cO_{X}(0,0))  {}= (1,0,0,0,0) , \\
    & h^\bullet(X;\cO_{X}(-16,-8)){}= (0,0,0,0,105) . 
  \eal 
\eeq 
Note already this extra contribution $h^3(X;\cO_{X}(-8,-5))=3$, which we will discuss in a moment. Analogous to tracing through the induced long exact sequences in \eqref{eq_hirzebruchseqex}, one can now determine 
\beq\label{eq_quinticexres1} 
  \bal 
    &h^\bullet(D;\cO_D(-1,0)) {}= (0,0,0,2) ,\quad &&{} h^\bullet(D;\cO_D(-2,-1)){}=(0,0,0,6) , \\
    &h^\bullet(D;\cO_D(0,-1)) {}= (0,0,3,1) ,\quad &&{} h^\bullet(D;\cO_D(0,0)){}= (1,0,0,1) , \\
    &h^\bullet(D;\cO_D(-8,-4)){}= (0,0,0,104). 
  \eal 
\eeq 
Likewise, we use those dimensions to determine the cohomology of the auxiliary bundle $\cE_D^*$ from the first sequence of \eqref{eq_quinticseqex}, yielding \beq\label{eq_quinticexres2} h^\bullet(D;\cE_D^*) = (0,2,3,21). \eeq In order to compute the cohomology of $\Omega^1_D$ from the second and final sequence in \eqref{eq_quinticseqex} some small additional input is required. Using \eqref{eq_quinticexres1} and \eqref{eq_quinticexres2}, the induced long exact sequence takes the form 
\beq\label{hannover96}
  \parbox{1cm}{\xymatrix{0 \ar[r] & \IR^0 = 0 \ar[r] & 0   \ar[r] & H^0(D;\Omega^1_D) \ar`[rd]`[l]`[dlll]`[d][dll] & \\
                      &  0 \ar[r] & \IR^2 \ar[r] & H^1(D;\Omega^1_D) \ar`[rd]`[l]`[dlll]`[d][dll] & \\
                      &  0 \ar[r] & \IR^3   \ar[r] & H^2(D;\Omega^1_D) \ar`[rd]`[l]`[dlll]`[d][dll] & \\
                      & \IR^{104} \ar[r] & \IR^{21}  \ar[r] & H^3(D;\Omega^1_D) \ar[r]
                      & 0.}}
\eeq
Whereas $H^0(D;\Omega^1_D) = 0$ and $H^1(D;\Omega^1_D) = 2$ follow immediately from the sequence, the remaining two cohomology groups seem to be ambiguous. However, one should keep in mind that via the symmetries in the Hodge diamond it follows that 
\beq
  H^3(D;\Omega^1_D) \cong H^{0,2}(D) \cong H^2(D;\cO_D) = 0. 
\eeq 
The remaining part of the sequence therefore reads 
\beq 
  0 \fto \IR^3 \injto H^2(D;\Omega^1_D) \fto \IR^{104} \surjto \IR^{21} \fto 0, 
\eeq 
such that via $3-\dim H^2(D;\Omega^1_D)+104-21=0$ as required for exactness we can determine the result 
\beq 
  h^\bullet(\Omega^1_D) = (0,2,86,0).
\eeq
This ultimately gives us the Hodge diamond 
\beq
  \begin{array}{ccccccc}
      &         &                 & h^{0,0} &                 &         &          \\
      &         & \mclap{h^{1,0}} &         & \mclap{h^{0,1}} &         &          \\
      & h^{2,0} &                 & h^{1,1} &                 & h^{0,2} &          \\
      \mclap{h^{3,0}} &         & \mclap{h^{2,1}} &         & \mclap{h^{1,2}} &         & \mclap{h^{0,3}}  \\
      & h^{2,0} &                 & h^{1,1} &                 & h^{0,2} &          \\
      &         & \mclap{h^{1,0}} &         & \mclap{h^{0,1}} &         &          \\
      &         &                 & h^{0,0} &                 &         &          
  \end{array}
  \qquad=\quad
  \begin{array}{ccccccc|l}
          &   &   & 1 &   &   &   & b^0=1 \\
          &   & 0 &   & 0 &   &   & b^1=0 \\
          & 0 &   & 2 &   & 0 &   & b^2=2 \\
      1   &   &86 &   &86 &   & 1 & b^3=174 \\
          & 0 &   & 2 &   & 0 &   & b^4=2 \\
          &   & 0 &   & 0 &   &   & b^5=0 \\
          &   &   & 1 &   &   &   & b^6=1 
  \end{array}
\eeq
for the octic Calabi-Yau 3-fold hypersurface $\IP^4_{11222}[8]$.

In prospect of section~\ref{sec_genbatyrev} we introduce\footnote{In order to avoid any confusion, note that the $\fhh^{p,q}_i$ here are entirely unrelated to the multiplicity factors $\fh^i(\cQ)$ of section~\ref{sec_basicalgorithm}.} the numbers $\fhh^{p,q}_i$ which refer to the contribution from the $i$th line bundle cohomology group $H^i(X;\cO_X(m,n))$ to the Hodge number $h^{p,q}$:
\beq\label{eq_weirdcontributionnumbers}
  h^i(X;\cO_X(m,n)) \quad \leadsto \quad \fhh^{p,q}_i
\eeq
It is clear from \eqref{hannover96} that $H^{2,1}(D)=H^2(D;\Omega^1_D)$ receives two different kinds of contributions, $\fhh_4^{21}(D)=83$ come from elements in $H^4(X,{\cal O}_X(m,n))$ and $\fhh^{21}_3(D)=3$ from elements in $H^3(X,\cO_X(m,n))$. Let us mention that this split can also be seen in the corresponding Gepner and Landau-Ginzburg orbifold models, where precisely 3~massless matter states come from so-called twisted sectors and the remaining 83 from the untwisted sector. As we will see in section \ref{sec_genbatyrev}, these states are also related to the non-geometric contributions in the Batyrev formula.

\subsection{Complete intersection subvarieties}\label{subsection:Complete intersection (CI) subvarieties}
In the last subsection we described a method to calculate the dimensions of the cohomology groups of the tangent bundle as well as the Hodge diamond for hypersurfaces up to three dimensions. Now we want to generalize these methods to the case where the subvariety does not arise as a hypersurface of a toric variety, but rather as a complete intersection of several hypersurface conditions. Here we will follow a similar path as before.

Let $\{S_1,\dots,S_l\}$ be a set of divisors on a toric variety $X$ such that their complete intersection subvariety is denoted by $S$. The tangent bundle of $S$ is then, in analogy to \eqref{eq_tangentbundlehypersurface}, given by the cohomology of the complex
\beq\label{eq_tangentbundlecicy}
  0 \fto \cO_S^{\oplus r} \stackrel{\alpha}{\injto} \bigoplus_{k=1}^n \cO_S(D_k) \stackrel{\beta}{\fto} \bigoplus_{i=1}^{l} \cO_S(S_i)\,,
\eeq
where as before the $D_k$ denote the vanishing divisors of the coordinates. As before we can perform a splitting of this complex into the two exact sequences 
\beq
  \bal
    &0 \fto \cO_S^{\oplus r} \injto \bigoplus_{k=1}^n \cO_S(D_k) \surjto \cE_S \fto 0 \\
    &0 \fto T_S \injto \cE_S \surjto \bigoplus_{i=1}^{l} \cO_S(S_i) \fto 0\,.
  \eal
\eeq
In order to calculate the Hodge diamond of the complete intersection $S$ we need the dual sequences which are given by
\beq\label{eq_dualcicotangentbundle}
  \bal
    & 0 \fto \cE_S^* \injto \bigoplus_{k=1}^n \cO_S(-D_k) \surjto \cO_S^{\oplus r} \fto 0 \\
    & 0 \fto \bigoplus_{i=1}^{l} \cO_S(-S_i) \injto \cE_S^* \surjto \Omega^1_S \fto 0.
  \eal
\eeq
Note that for $l=1$ this precisely reproduces the hypersurface result \eqref{eq_splittangentbundle}.

For hypersurfaces we were able to use the Koszul sequence \eqref{eq_koszulsequence} in order to calculate the cohomologies of line bundles over the hypersurface. Here the situation is a bit more involved and we have to employ the generalized Koszul sequence
\beq
  \parbox{1cm}{
  \xymatrix{ 
      & & \mllap{0 \fto
        \smash{\cO_X\bigg({\textstyle-\sum\limits_{j=1}^{l} S_{j}}\bigg)}} \ar[r] &
        \smash{\bigoplus\limits_{\mclap{i_1<\ldots<i_{l-1}}} \cO_X\bigg({\textstyle -\sum\limits_{j=1}^{l-1} S_{i_j}}\bigg)} \ar`[rd]`[l]`[dlll]`[d][dll] & 
        \\ 
    & 
        \smash{\bigoplus\limits_{\mclap{i_1<\ldots<i_{l-2}}}\cO_X\bigg({\textstyle-\sum\limits_{j=1}^{l-2} S_{i_j}}\bigg)} \ar[r] &
    \ldots\big. \ar[r] & 
        \smash{\bigoplus\limits_{\mclap{i_1<i_2}} \cO_X\left(-S_{i_1}-S_{i_2}\right)} \ar`[rd]`[l]`[dlll]`[d][dll] & 
        \\ & 
        \smash{\bigoplus\limits_{i_1} \cO_X(-S_{i_1})} \ar[r] & 
        \cO_X  \ar[r] & 
        \cO_S \fto 0\,. & 
  }}^\big._\big.
\eeq
Note that given $\Lambda^1=\bigoplus_{i_1}\cO_X(-S_{i_1})$ all line bundles prior in the sequence chain can be interpreted as higher exterior powers. Twisting the whole sequence by $\cO_X(D)$ leads to
\begin{small}
    \beq\label{eq_generalizedkoszulwithdiviso}
      \parbox{1cm}{\xymatrix{ 
          & 
            \hspace{-1.3cm}0 \fto \smash{\cO_X\bigg({\textstyle-\sum\limits_{j=1}^l S_{j}+D}\bigg)} \ar[r] & 
            \ldots \ar[r] & 
            \smash{\bigoplus\limits_{\mclap{i_1<i_2}} \cO_X\left(-S_{i_1}-S_{i_2}+D\right)} \ar`[rd]`[l]`[dlll]`[d][dll] & 
            \\ & 
            \smash{\bigoplus\limits_{i_1} \cO_X(-S_{i_1}+D)} \ar[r] & 
            \cO_X(D)\big. \ar[r] & 
            \cO_S(D) \fto 0\,. &  
        }}^\big._\big.
    \eeq
\end{small}\noindent
In contrast to the situation with a simple hypersurface, we are not finished yet, since the sequence \eqref{eq_generalizedkoszulwithdiviso} is not a short exact one and hence does not give rise to a long exact sequence in cohomology. But one can easily see that an exact sequence of length $l+2$ yields $l$ short exact sequences using several auxiliary sheaves $\cI_k$:
\beq\label{eq_generalizedkoszulsplitted}
\boxed{
    \bal
        0\fto \cO_X\bigg({\textstyle -\sum\limits_{j=1}^l S_{j}+D}\bigg) \injto
        & {} \bigoplus_{\mclap{i_1<\ldots<i_{l-1}}} \cO_X\bigg({\textstyle -\sum\limits_{j=1}^{l-1}S_{i_j}+D}\bigg) \surjto \cI_1 \fto 0 \\
        0\fto\cI_1 \injto 
        & {} \bigoplus_{\mclap{i_1<\ldots<i_{l-2}}} \cO_X\bigg({\textstyle-\sum\limits_{j=1}^{l-2}S_{i_j}+D}\bigg) \surjto \cI_2 \fto 0 \\
        & {}\quad \vdots \\
        0\fto\cI_{l-2} \injto 
        & {} \bigoplus_{i_1}\cO_X\left(-S_{i_1}+D\right) \surjto \cI_{l-1} \fto 0 \\
        0\fto\cI_{l-1} \injto
        & {} \cO_X(D) \surjto \cO_S(D) \fto 0
    \eal}
\eeq
These are the ones we are actually going to use in explicit calculations. This means that in order to derive the dimensions of the cohomology groups of $\cO_S(D)$ we first have to write down all the required long exact sequences and derive the cohomologies of $l-1$ auxiliary sheafs.

For instance, for a complete intersection $S$ of two hypersurfaces $S_1$ and $S_2$, the splitting of the generalized Koszul sequence \eqref{eq_generalizedkoszulsplitted} is given by
\begin{small}
    \beq\label{eq_generalizedkoszulsplittedtwohypersurfaces}
        \bal
            & 0 \fto \cO_X\left(-S_1-S_2+D\right)  \injto \cO_X\left(-S_1+D\right)\oplus\cO_X\left(-S_2+D\right) \surjto \cI_1  \fto 0 \\
            & 0 \fto \cI_1 \injto \cO_X(D) \surjto \cO_S(D) \fto 0 
        \eal
    \eeq
\end{small}
which is sufficient to calculate all necessary ingredients of \eqref{eq_dualcicotangentbundle}. Naturally, the ability to make use of exactness is once again critical for actual computations.

\subsubsection*{\texorpdfstring{Example: Hodge diamond of the $\IP_{111122}[4,4]$ CICY}{Example: Hodge diamond of the P111122[4,4] CICY}}\label{sec_P111122CICY}
Let us now see how the method of calculating the Hodge diamond for a CICY works in detail for the case of two intersecting hypersurfaces, by examining the specific case of $S:=\mathbb P_{111122}^5[4,4]$ living in the weighted projective ambient space $X:=\IP_{111122}^5$. The corresponding toric data is given in table \ref{tab_P_111122[4,4]}.

\begin{table}[ht]
  \centering
	\begin{tabular}{r@{\,$=$\,(\,}r@{,\;\;}r@{,\;\;}r@{,\;\;}r@{,\;\;}r@{\,)\;\;}|c|c|c} 
		\multicolumn{6}{c|}{vertices of the} & coords & GLSM charges & {divisor class}${}^\big.$ \\
		\multicolumn{6}{c|}{polyhedron / fan}      &        & $Q^1$ & \\ 
					\hline\hline
		$\nu_1$ & $-1$ & $-1$ & $-1$ & $-2$  & $-2$& $u_1$ & 1 & $H$\uspc \\
		$\nu_2$ &   1  &   0  &   0  &   0   &  0  & $u_2$ & 1 & $H$ \\
		$\nu_3$ &   0  &   1  &   0  &   0   &  0  & $u_3$ & 1 & $H$ \\
		$\nu_4$ &   0  &   0  &   1  &   0   &  0  & $u_4$ & 1 & $H$ \\
		$\nu_5$ &   0  &   0  &   0  &   1   &  0  & $u_5$ & 2 & $2H$ \\
		$\nu_6$ &   0  &   0  &   0  &   0   &  1  & $u_6$ & 2 & $2H$\lspc \\
					\hline
					\multicolumn{7}{c|}{conditions:}            & 4 & \uspc \\
					\multicolumn{7}{c|}{}                       & 4 &
	\end{tabular}
	\\[2mm] $\text{intersection form:}\quad \frac{1}{4} H^5$
	\\[2mm] ${\rm SR}(X) = \langle u_1 u_2 u_3 u_4 u_5 u_6 \rangle$
  \caption{\small Toric data for the CICY threefold $S:=\IP_{111122}^5[4,4]$ living in the ambient space $X:=\IP_{111122}^5$.}
  \label{tab_P_111122[4,4]}
\end{table}

\vspace{0.3cm}
\noindent
For this example the sequences \eqref{eq_dualcicotangentbundle} reduce to
\beq\label{eq_dualcicotangentbundleP_111122[4,4]}
	\bal
		& 0 \fto \cE_S^* \injto \cO_S(-1)^{\oplus 4} \oplus  \cO_S(-2)^{\oplus 2} \surjto \cO_S \fto 0 \\
		& 0 \fto \cO_S(-4) \oplus \cO_S(-4) \injto \cE_S^* \surjto \Omega_S \fto 0
	\eal
\eeq
and hence we need to determine the cohomologies of the line bundles $\cO_S(-1)$, $\cO_S(-2)$, $\cO_S$, $\cO_S(-4)$ over the CICY. This can be done by employing equations \eqref{eq_generalizedkoszulsplittedtwohypersurfaces} which give us the four pairs of equations, one pair for each line bundle
\begin{subequations}
	\label{eq_sequenceshodgediamondP111122}
	\begin{eqnarray}\label{eq_sequenceshodgediamondP111122a}
		\bal
			0 & \fto \cO_{X}(-9)  \injto  \cO_{X}(-5)^{\oplus2}  \surjto  \cI_{a}  \fto  0 \\
			0 & \fto \cI_{a}  \injto  \cO_{X}(-1)  \surjto  \cO_{S}(-1)  \fto  0 
		\eal
	\end{eqnarray}
	\begin{eqnarray}\label{eq_sequenceshodgediamondP111122b}
		\bal
			0 & \fto \cO_{X}(-10)  \injto  \cO_{X}(-6)^{\oplus2}  \surjto  \cI_{b}  \fto  0 \\
			0 & \fto \cI_{b}  \injto  \cO_{X}(-2)  \surjto  \cO_{S}(-2)  \fto  0 
		\eal
	\end{eqnarray}
	\begin{eqnarray}\label{eq_sequenceshodgediamondP111122c}
		\bal
			0 & \fto \cO_{X}(-8) \injto  \cO_{X}(-4)^{\oplus2}  \surjto  \cI_{c}  \fto  0 \\
			0 & \fto \cI_{c}  \injto  \cO_{X}  \surjto  \cO_{S}  \fto  0 
		\eal
	\end{eqnarray}
	\begin{eqnarray}\label{eq_sequenceshodgediamondP111122d}
		\bal
			0 & \fto \cO_{X}(-12) \injto  \cO_{X}(-8)^{\oplus2}  \surjto  \cI_{d}  \fto  0 \\
			0 & \fto \cI_{d}  \injto  \cO_{X}(-4)  \surjto  \cO_{S}(-4)  \fto  0 
		\eal
	\end{eqnarray}
\end{subequations}
Deriving the corresponding long exact sequences of the cohomology groups allows us to determine for each pair first the dimensions of the cohomologies of the auxiliary sheaf and then in the second step the one for the line bundle itself. The computation of the dimensions of the cohomology groups of the line bundles is easily done using our algorithm \cite{cohomCalg:Implementation}. For some of the line bundles in \eqref{eq_sequenceshodgediamondP111122} all cohomology groups vanish. For those where this is not the case we find 
\beq
	\bal
		& h^\bullet(X;\cO_{X}(-9)) {}= (0,0,0,0,0,4),~ 
		& h^\bullet(X;\cO_{X}(-10)){}= (0,0,0,0,0,12),\\
		& h^\bullet(X;\cO_{X}(-8)) {}= (0,0,0,0,0,1),~
		& h^\bullet(X;\cO_{X}(-12)){}= (0,0,0,0,0,58)\\
		& h^\bullet(X;\cO_{X})     {}= (1,0,0,0,0,0)
	\eal
\eeq
from which follows the cohomology of the auxiliary sheafs / bundles
\beq
	\bal
		h^\bullet(X;\cI_a) &{}= (0,0,0,0,4,0),\quad
		h^\bullet(X;\cI_b) {}= (0,0,0,0,12,0), \\
		h^\bullet(X;\cI_c) &{}= (0,0,0,0,1,0), \quad
		h^\bullet(X;\cI_d) {}= (0,0,0,0,56,0)\,.
	\eal
\eeq
Taking this into account one can use the second sequences from \eqref{eq_sequenceshodgediamondP111122a}-\eqref{eq_sequenceshodgediamondP111122b} to read off
\beq\label{eq_cohomslinebundleP_111122[2,2]}
	\bal
		& h^\bullet(S;\cO_{S}(-1)) {}= (0,0,0,4),\quad
		& h^\bullet(S;\cO_{S}(-2)) {}= (0,0,0,12) \\
		& h^\bullet(S;\cO_{S})     {}= (1,0,0,1),\quad 
		& h^\bullet(S;\cO_{S}(-4)) {}= (0,0,0,56)\,,
	\eal
\eeq
where $h^\bullet(S;\cO_{S})$ already presents the expected first row of the Hodge diamond. Now we can proceed in the same way as we did in the last subsection and plug this into the first equation of \eqref{eq_dualcicotangentbundleP_111122[4,4]} to get
\beq
  h^\bullet(S;\cE_S^*) = (0,1,0,39)\,.
\eeq
We insert this result together with $h^\bullet(S;\cO_{S}(-4))$ from equations \eqref{eq_cohomslinebundleP_111122[2,2]} into the second equation in \eqref{eq_dualcicotangentbundleP_111122[4,4]}. In order to derive a unique result from the long exact sequence, we have to use the fact that the complete intersection is Calabi-Yau which implies that $h^0(S;\Omega^1_S) = 0$ and find
\beq
  h^\bullet(S;\Omega^1_S)  {}= (0,1,73,0)
\eeq
Since this is the second row of the Hodge diamond we are looking for and since $S$ is Calabi-Yau, we can write down the full Hodge diamond of $\IP_{111122}[4,4]$:
\beq
	\begin{array}{ccccccc|l}
				&   &   & 1 &   &   &   & b^0=1 \\
				&   & 0 &   & 0 &   &   & b^1=0 \\
				& 0 &   & 1 &   & 0 &   & b^2=1 \\
		1 &   &73 &   &73 &   & 1 & b^3=148 \\
				& 0 &   & 1 &   & 0 &   & b^4=1 \\
				&   & 0 &   & 0 &   &   & b^5=0 \\
				&   &   & 1 &   &   &   & b^6=1 
	\end{array}
\eeq
Note again that by no means we are using any properties special to this geometry, i.e.~the described procedure is completely algorithmic and can be analogously applied to any other setting as long as enough zeros appear in the cohomologies to make use of exactness. All the laborious and somewhat confusing steps involved in this computation can be easily carried out with the Koszul module of \cite{cohomCalg:Implementation}, which automates precisely the steps outlined above.

\subsection[Hodge numbers of 4- and 5-dimensional toric subspaces]{Hodge numbers of 4- and 5-dimensional \\ toric subspaces}
Naturally one would like to extend the computation of the Hodge diamond to higher-dimensional subspaces, which continues the discussion in sec.~\ref{sec_highdimambienthodgenumbers}.

\subsubsection*{Hypersurfaces}
For hypersurfaces the same method is applied, albeit a lot more sequences are involved and many more variations lead to the same result of $\Omega^2_D$, where $D$ is the hypersurface in question. Therefore we will only provide a single example of deriving the cohomology of this sheaf / bundle for hypersurfaces. 

The general idea is to apply the aforementioned methods of sec.~\ref{sec_highdimambienthodgenumbers} to \eqref{eq_dualhypersurfacetangent}. Using the simplified sequence \eqref{eq_lambdatwolinebdls}, where $A$ is a line bundle, the bottom sequence of \eqref{eq_dualhypersurfacetangent} yields
\beq
  0 \fto \Omega^1_D \otimes \cO_D(-D) \injto \Lambda^2\cE_D^* \surjto \Omega^2_D \fto 0,
\eeq
such that it remains to determine the cohomology of the bundles on the right and in the middle. The cohomology of $\Omega^1_D\otimes\cO_D(-D)$ can be obtained by simply tensoring the dualized Euler sequences of \eqref{eq_dualhypersurfacetangent} with $\cO_D(-D)$, which yields
\beq
  \bal
      & 0 \fto \cE_D^* \otimes \cO_D(-D) \injto \bigoplus_{k=1}^n \cO_D(-D-D_k) \surjto \cO_D(-D)^{\oplus r} \fto 0 \\
        & 0 \fto \cO_D(-2D) \injto \cE_D^*\otimes\cO_D(-D) \surjto \Omega_D^1\otimes\cO_D(-D) \fto 0.
    \eal
\eeq
The second ingredient $\Lambda^2\cE_D^*$ is determined from \eqref{eq_lambdatwovarA} and the middle sequence of \eqref{eq_dualhypersurfacetangent}, i.e.~we have to consider
\beq
  \bal
      & 0 \fto Q_1 \injto \bigoplus_{i<j} \cO_D(-D_i-D_j) \surjto \cO_D^{\oplus\binom{r}{2}} \fto 0 \\
        & 0 \fto \Lambda^2 \cE_D^* \injto Q_1 \surjto (\cE_D^*)^{\oplus r} \fto 0.
    \eal
\eeq

As mentioned before, running through the sequences relies on the ability to make use of the exactness of the induced long exact sequence. Therefore it might be necessary to use another ``way'' through the sequences to complete actual computations, see section~\ref{sec_highdimambienthodgenumbers} again.

\subsubsection*{Complete Intersection}
As before, the generalization from a hypersurface to the case of a complete intersection of $l$ hypersurfaces is straightforward. Instead of using equation \eqref{eq_dualhypersurfacetangent} we use the second of  equation \eqref{eq_dualcicotangentbundle}. Together with equation \eqref{eq_lambdatwolinebdls} we end up with

\beq\label{eq_Lamda2CI}
  0 \fto \Omega^1_S \otimes \bigoplus_{j=1}^l \cO_S(-S_j) \injto \Lambda^2\cE_S^* \surjto \Omega^2_S \fto 0.
\eeq
Following the same procedure as for the hypersurface case, we can determine the first part of \eqref{eq_Lamda2CI} by
\beq
  \bal
      & 0 \fto \cE_S^* \otimes \bigoplus_{j=1}^l\cO_S(-S_j) \injto \bigoplus_{j=1}^l \bigoplus_{k=1}^n \cO_S(-S_j-D_k) \surjto \bigoplus_{j=1}^l \cO_S(-S_j)^{\oplus r} \fto 0 \\
        & 0 \fto  \bigoplus_{j=1}^l \bigoplus_{i=1}^l\cO_S(-S_j-S_i) \injto \cE_S^*\otimes  \bigoplus_{j=1}^l\cO_S(-S_j) \surjto \Omega_S^1\otimes \bigoplus_{j=1}^l\cO_S(-S_j) \fto 0.
    \eal
\eeq
The second part $\Lambda^2\cE_S^*$ is again determined from \eqref{eq_lambdatwovarA} and the second sequence of \eqref{eq_dualcicotangentbundle}, i.e.~
\beq
  \bal
      & 0 \fto Q_1 \injto \bigoplus_{i<j} \cO_S(-D_i-D_j) \surjto \cO_S^{\oplus\binom{r}{2}} \fto 0 \\
        & 0 \fto \Lambda^2 \cE_S^* \injto Q_1 \surjto (\cE_S^*)^{\oplus r} \fto 0.
    \eal
\eeq
Obviously, the complexity and number of steps involved in a full computation rapidly increases with the number of intersections and the number of dimensions. Thanks to the algorithmic nature of our approach, however, the entire process has been automated in the {\cohomCalg} Koszul module which operates precisely on the procedure outlined here.

\section{Cohomology for Orientifolds and Orbifolds}\label{sec_orientifolds}
In order to  reduce  the $\cN$=2 space-time supersymmetry of  Type IIA/B  superstring theories on Calabi-Yau 3-folds down to $\cN$=1, one needs to consider orientifolds. Often, only the ingredients invariant under this symmetry survive the subsequent orientifold projection, such that the theory actually lives on the quotient  space. For matter zero modes, using the usual splitting into the eigenvalues of this $\IZ_2$-action, it is necessary to consider the invariant and anti-invariant parts of the cohomology groups we have been discussing so far. 

A second important application of equivariant cohomology is found in orbifold  constructions often performed  in heterotic string compactifications. Since in naive Calabi-Yau three-fold compactifications quantities like the Euler characteristic are directly tied to physical properties like e.g.~the number of matter generations, orbifold constructions are often used to build spaces with suitable topological numbers. Usually one finds an abundance of ``plain'' spaces with huge topological invariants, whereas the phenomenologically interesting areas of the topological moduli space are sparsely populated. Orbifolds  can greatly help in this aspect. For example, letting $\IZ_5$ act freely on the quintic Calabi-Yau 3-fold shows $\chi(\IC\IP^4[5]/\IZ_5) = \frac{1}{5} \chi(\IC\IP^4[5])$.

The goal of this section is therefore to develop tools how equivariant cohomology groups can be computed. Since our algorithm provides explicit representatives for the cohomological elements, it is tailor-made for this purpose. The main question is, how these extra multiplicity factors $\fh_i(Q)$ in \eqref{vflwolfburg} contribute. We will present two conjectures for the simple computation of the dimensions of equivariant cohomology groups with line bundles on toric spaces, where the first one deals with the simpler $\mathbb Z_2$ case and the second one with the generalization to any finite group. First, in order to have a non-trivial cross-check we utilize a number of topological tools.

\subsection{\texorpdfstring{Topological invariants for $\IZ_2$ involutions}{Topological invariants for Z2 involutions}}\label{sec_Z2invariants}
A very useful tool in complex geometry is the Riemann-Roch-Hirzebruch theorem. Given a holomorphic vector bundle $V$ on some complex manifold $X$ of dimension $n$, it allows to compute the Euler characteristic of this bundle via its Chern character and the Todd class of the base manifold, i.e.
\beq\label{eq_riemannrochhirzebruch}
  \chi(X;V) \ce \sum_{i=0}^n (-1)^i\dim H^i(X;V) \stackrel{\text{RRH}}{=} \int_X {\rm ch}(V) \, {\rm Td}(X),
\eeq
where ${\rm ch}(V)$ refers to the Chern character of $V$, a polynomial expression of the Chern classes
\beq
  \bal
    {\rm ch}(V) = \dim(V) {}&{}+ c_1(V) + \frac{c_1(V)^2 - c_2(V)}{2} \\
		 {}&{}+ \frac{c_1(V)^3 - 3c_1(V)c_2(V) + 3c_3(V)}{6} + \dots,
	\eal
\eeq
satisfying ${\rm ch}(V\oplus W)={\rm ch}(V)+{\rm ch}(W)$ as well as ${\rm ch}(V\otimes W)={\rm ch}(V)\,{\rm ch}(W)$ and ${\rm Td}(X)={\rm Td}({\rm T}_X)$ is the Todd class of the base space's tangent bundle, which can for a holomorphic vector bundle also be represented by a Chern class polynomial
\beq
  {\rm Td}(E) = 1+\frac{1}{2}c_1(E) + \frac{1}{12}\Big(c_1(E)^2 + c_2(E)\Big)
  + \dots \; .
\eeq
Note that for line bundles the Chern character simplifies to the simple Taylor expansion
\beq
  {\rm ch}(L) = {\rm e}^{c_1(L)} = \sum_m \frac{c_1(L)^m}{m!} = 1 + c_1(L) + \frac{c_1(L)^2}{2} + \dots
\eeq
that naturally truncates at the dimension of the base space, leaving only a finite number of non-zero terms in the sum.

Naturally, one would like to extend the index formula \eqref{eq_riemannrochhirzebruch} in some way to settings subject to a symmetry action on the base space, e.g.~the $\IZ_2$ space-time involution $\Omega\sigma$ of a typical orientifold operation. The vector bundle $V$ must be compatible with the $\IZ_2$ action $\sigma$ of the orientifold involution, i.e.~we require the induced mapping $\sigma^*$ to fulfill $\pi\circ\sigma^* = \sigma$ where $\pi:V\fto X$ is the bundle's projection mapping. Then $\sigma$ induces the splitting
\beq\label{eq_orientifoldgroupsplitting}
  H^i(X;V) = H^i_+(X;V) \oplus H^i_-(X;V)
\eeq
of the cohomology groups. Following a general theorem, the Euler characteristic of the orientifold's ``downstairs'' quotient space $X/\sigma$, i.e.~the invariant part of the splitting, can be expressed as
\beq\label{eq_Z2cosetformula}
  \chi(X/\sigma;\tilde V) = \chi_+(X;V) = \sum_{i=0}^n (-1)^i h^i_+(X;V) = \frac{\chi^e(X;V) + \chi^\sigma(X;V)}{2},
\eeq
where $\chi^e$ and $\chi^\sigma$ are Euler characteristica associated to the two group elements of $\IZ_2=\{e,\sigma\}$. Here $\tilde V$ corresponds to the bundle $V$ on the quotient space $X/\sigma$. Since $e$ is the unit element, $\chi^e$ actually corresponds to the ordinary Euler characteristic
\beq\label{eq_normalsplitting}
  \bal
    \chi^e(X;V) & {}= \chi(X;V) \ce \sum_{i=0}^n (-1)^i h^i(X;V) \\
		&{}= \sum_{i=0}^n (-1)^i \Big( \dim H^i_+(X;V) + \dim H^i_-(X;V) \Big),
	\eal
\eeq
and from the splitting on the right hand side of this equation one directly obtains
\beq\label{eq_holLefschetzNumber}
  \chi^\sigma(X;V) = \sum_{i=0}^n (-1)^i \Big( \dim H^i_+(X;V) - \dim H^i_-(X;V) \Big),
\eeq
which gives us a sort of measure for the dimensional asymmetry of the splitting. This quantity is called the holomorphic Lefschetz number and is related to the fixpoint set of $\sigma$, i.e.~to the so-called O-planes in an orientifold setting. The simple split of the cohomology groups also allows to provide the Euler characteristic of the anti-invariant part. From \eqref{eq_normalsplitting} and \eqref{eq_holLefschetzNumber} it follows
\beq
  \chi_-(X;V) = \sum_{i=0}^n (-1)^i h^i_-(X;V) = \frac{\chi^e(X;V) - \chi^\sigma(X;V)}{2}
\eeq
in obvious similarity to \eqref{eq_Z2cosetformula}. Both $\chi_+(X;V)$ and $\chi_-(X;V)$ are used as highly nontrivial checks for the computations carried out in the next section.

Analogous to the Riemann-Roch-Hirzebruch theorem \eqref{eq_riemannrochhirzebruch} the holomorphic Lefschetz theorem and the Atiyah-Bott theorem allow to compute the Lefschetz number via an index formula
\beq\label{eq_holLefschetztheorem}
  \chi^\sigma(X;V) = \int_{X^\sigma} {\rm ch}_\sigma(V)\,\frac{{\rm Td}(T_{X^\sigma})}{{\rm ch}_\sigma\big(\Lambda_{-1}(\bar{N}_{X^\sigma})\big)},
\eeq
which---as mentioned before---only depends on the fixpoint set of the involution $\sigma$. In this expression the $\Lambda_{-1}(\bar{N}_{X^\sigma})$ refers to the formal alternating sum of the exterior powers of the complex conjugate normal bundle of the orientifold involution fixpoint set $X^\sigma\subset X$, i.e.
\beq
  \Lambda_{-1}(\bar{N}_{X^\sigma}) = \sum_{i=0}^n (-1)^i \Lambda^i(\bar{N}_{X^\sigma}).
\eeq
In order to define the equivariant Chern character ${\rm ch}_\sigma(V)$, the vector bundle $V$ is first decomposed into a direct sum of $\sigma_*$-eigenbundles, i.e.~bundles $V_k$ which are either invariant or anti-invariant under the induced $\sigma_*$-action. 

One of the main simplifications for $\IZ_2$-involutions derives from the fact that the induced action on the normal bundle is simply
\beq
  \sigma_*(N_{\rm{X^\sigma}}) = -N_{\rm{X^\sigma}}.
\eeq
For the vector bundle $V$ one first decomposes it into a direct sum of eigenbundles $V=V_1\oplus\cdots\oplus V_m$ with eigenvalues $\rho_k= \pm 1$ and then defines
\beq
  {\rm ch}_\sigma(V) \ce \sum_{k=1}^m \rho_k \,{\rm ch}(V_k).
\eeq
Some further information on these definitions can be found in the appendix of \cite{Blumenhagen:2010ja} and references therein. It should be noted that the holomorphic Lefschetz theorem can be regarded as a special case of the Atiyah-Singer fixed point theorem and the index formula is also referred to as the Atiyah-Bott theorem, see \S17 of \cite{Shanahan:Book1978}.

\subsection{\texorpdfstring{An algorithm conjecture for $\IZ_2$-equivariance}{An algorithm conjecture for Z2-equivariance}}\label{sec_Z2algconjecture}
The algorithm for the computation of line bundle cohomologies on toric varieties \cite{Blumenhagen:2010pv} provides actual representatives for the cohomology group generators in the form of so-called rationoms, i.e.~rational functions with a single monomial in the numerator and denominator, as long as only trivial multiplicities for the individual monomials are involved.  Consider for example the projective sphere $\IC\IP^3$ and the ``sign flip'' involution
\beq
  \sigma:(x_1,x_2,x_3,x_4)\mapsto(-x_1,x_2,x_3,x_4)
\eeq
on the homogeneous coordinates of the base, which due to the projective equivalences is equivalent to the involution
\beq
  \tau:(x_1,x_2,x_3,x_4)\mapsto(x_1,-x_2,-x_3,-x_4).
\eeq
The fixpoint set of this involution therefore consist of two components: The divisor $\{u_1=0\}\cong\IC\IP^2$ and the isolated fixpoint $(1,0,0,0)$. Due to the simplicity of the Stanley-Reisner ideal 
\beq
  {\rm SR}(\IC\IP^3) = \langle u_1 u_2 u_3 u_4 \rangle,
\eeq
the contributing rationoms for the computation of $h^*(\IC\IP^3;\cO(k))$ are of a particularly simple form:
\beq\label{eq_CP3bundleMonomials}
  \bal
    & \text{for $k\ge0$: }   && \left\{ x_1^a x_2^b x_3^c x_4^d : a+b+c+d=k \right\}, \\
	  & \text{for $k\le-4$:} && \left\{ \frac{1}{x_1^{a+1} x_2^{b+1} x_3^{c+1} x_4^{d+1}} : a+b+c+d=-k-4 \right\}.
	\eal 
\eeq
In order to identify the overall sign each rationom picks up, one can simply apply the involution $\sigma$ to it. However, consider for example the bundle $\cO(-5)$ and the corresponding sign under the involutions $\sigma$ and $\tau$:
\beq\label{eq:EquivariantCountingMismatch}
  \underbrace{\frac{1}{x_1^2 x_2 x_3 x_4}}_{\substack{\sigma\to+ \\ \tau\to-}}, 
	\underbrace{\frac{1}{x_1 x_2^2 x_3 x_4}}_{\substack{\sigma\to- \\ \tau\to+}}, 
	\underbrace{\frac{1}{x_1 x_2 x_3^2 x_4}}_{\substack{\sigma\to- \\ \tau\to+}}, 
	\underbrace{\frac{1}{x_1 x_2 x_3 x_4^2}}_{\substack{\sigma\to- \\ \tau\to+}}
	\quad\leadsto\quad
	\bal
	  & \sigma: && (1_+,3_-) \\
		& \tau:   && (3_+,1_-)
	\eal
\eeq
There is obviously a  mismatch in the counting of signs between the two equivalent involutions of the base, which can be seen in almost all bundles $\cO(k)$.

Ultimately, this is due to the naive application of the base involutions to the representatives of the bundle cohomology. In mathematical terms, one needs to uplift the $\IZ_2$-action on the base to an $\IZ_2$-action on the bundle $L=\cO(k)$, which is called an equivariant structure and makes the diagram
\beq
  \xymatrix{L \ar@{-->}[r]^{\phi_\sigma} \ar@{->>}[d]_\pi & L \ar@{->>}[d]^\pi \\
	          \IC\IP^3 \ar[r]^\sigma & \IC\IP^3}
	\qquad
\eeq
commutative. More precisely, for a generic group $G$, each element $g\in G$ induces a mapping $g:X\longrightarrow X$ on the base geometry and has a corresponding uplift $\phi_g:L\fto L$ compatible with the bundle structure. This uplift defines an equivariant structure, if it preserves the group structure, i.e.~if $\phi_g \circ \phi_h = \phi_{gh}$ such that the mapping is a group homomorphism.

The apparent inconsistency of \eqref{eq:EquivariantCountingMismatch} therefore stems from the false assumption that the equivalent involutions $\sigma$ and $\tau$ in the base geometry give rise to equivalent equivariant structures $\phi_\sigma$ and $\phi_\tau$ on the bundle $\cO(k)$. For such a setting it is therefore important to specify the equivariant structure, i.e.~the uplift of the base involution to the bundle, as well.

\begin{table}[t]
  \centering
    \begin{tabular}{r@{\,$=$\,(\,}r@{,\;\;}r@{\,)\;\;}|c|cccccc|c} 
      \multicolumn{3}{c|}{vertices of the} & coords & \multicolumn{6}{c|}{GLSM charges} & {divisor class}${}^\big.$ \\
      \multicolumn{3}{c|}{polyhedron / fan}      &        & $Q^m$ & $Q^n$ & $Q^p$ & $Q^q$ & $Q^r$ & $Q^s$ & \\ 
            \hline\hline
      $\nu_1$ & $-1$ & $-1$ & $u_1$ & 1 & 0 & 0 & 1 & 0 & 0 & $H$\uspc \\
      $\nu_2$ &   1  &   0  & $u_2$ & 1 & 0 & 1 & 0 & 1 & 0 &     \\
      $\nu_3$ &   0  &   1  & $u_3$ & 1 & 1 & 0 & 0 & 0 & 0 &     \\
      $\nu_4$ &   0  & $-1$ & $u_4$ & 0 & 1 & 0 & 0 & 1 & 0 & $E_a$ \\
      $\nu_5$ & $-1$ &   0  & $u_5$ & 0 & 0 & 1 & 0 & 0 & 0 & $E_b$ \\
			$\nu_6$ &   1  &   1  & $u_6$ & 0 & 0 & 0 & 1 & 0 & 0 & $E_c$ \\
			$\nu_7$ & $-1$ &   1  & $u_7$ & 0 & 0 & 0 & 0 & 1 & 1 & $E_d$ \\
      $\nu_8$ &   1  & $-1$ & $u_8$ & 0 & 0 & 0 & 0 & 0 & 1 & $E_e$\lspc \\ \hline
    \end{tabular}
  \\[2mm] $I_{\widetilde{dP_5}} = H E_a + H E_b - H^2 - E_e^2 - 2E_a^2 - 2E_b^2 + E_a E_e + E_b E_d - E_d^2 -E_c^2$
  \\[2mm] $\bal {\rm SR}(\widetilde{dP_5}) = \langle & u_1 u_2,\;  u_1 u_3,\;  u_1 u_6,\;  u_1 u_7,\;  u_1 u_8,\;  u_2 u_3,\;  u_2 u_4,\;   \\
	                                    & u_2 u_5,\;  u_2 u_7,\;  u_3 u_4,\;  u_3 u_5,\;  u_3 u_8,\;  u_4 u_5,\;  u_4 u_6,\;   \\
																			& u_4 u_7,\;  u_5 u_6,\;  u_5 u_8,\;  u_6 u_7,\;  u_6 u_8,\;  u_7 u_8 \rangle \eal$
  \caption{\small Toric data for the non-generic $\widetilde{dP_5}$ surface, which arises via two additional blowups from the standard $dP_3$ and differs from the standard $dP_5=\IP^5[2,2]$.}
  \label{tab_dP5}
\end{table}

A second non-trivial aspect in the computation of equivariant cohomology comes from the non-trivial multiplicities appearing for some denominator monomials of our algorithm. One could question, if the invariant and anti-invariant monomial contributions with non-trivial multiplicities might nevertheless contribute unconventionally to the invariant and anti-invariant cohomology groups. As a highly non-trivial check for this issue, we consider the non-standard del~Pezzo-5 surface, which has a toric description similar to $dP_1$, $dP_2$ and $dP_3$. The relevant toric data is summarized in table~\ref{tab_dP5}. Due to the high number of 20~Stanley-Reisner ideal generators, this example yields 200 potentially contributing monomial denominators, where 56 of these have multiplicity~2 and two have multiplicity~3. Now, consider the involution
\beq\label{eq:dP5signflip}
  \sigma:u_1\mapsto -u_1,
\eeq
which is equivalent to 64 different ``sign flips'' due to the projective equivalences. The fixpoint set in the base can be determined to be
\beq
  X^\sigma \ce {\rm FP}_\sigma(\widetilde{dP_5}) = \{ u_1 = 0 \} \cup \{ u_6 = 0 \} \cup \{ u_7 = 0 \} \cup \{ u_8 = 0 \},
\eeq
giving four non-intersecting $\IP^1$s inside the $\widetilde{dP_5}$. For the equivariant structure we use the canonical uplift of \eqref{eq:dP5signflip}. Via a proper computation of \eqref{eq_holLefschetztheorem} the resulting Lefschetz number is
\beq
  \bal
    \chi^\sigma(\widetilde{dP_5};\cO(m,\dots,s)) = {}&{} \left(\frac{1}{4} + \frac{-m + n + p}{2}\right) + (-1)^n\left(\frac{1}{4} + \frac{m - q}{2}\right) \\
		   &{}+ (-1)^{m+n+r}\left(\frac{1}{4} + \frac{n + p - r}{2} \right) \\
			 &{}+ (-1)^{m + n + r + s}\left( \frac{1}{4} + \frac{r - s}{2}\right),
	\eal
\eeq
which allows to check whether a multiplicity-3 rationom like $\frac{1}{u_1 u_6 u_7 u_8}$ or $\frac{1}{u_2 u_3 u_4 u_5}$ entirely contributes to the invariant or anti-invariant cohomology. Likewise, we checked an abundance of other examples. The empirical data therefore leads us to pose the following:

\begin{description}
  \item[Conjecture for $\IZ_2$-equivariant cohomology:]\label{Z2conjection} Given a toric space $X$, an involution on the base $\sigma:X\fto X$ as well as an equivariant structure, the lifted involution mapping can be directly applied to the rationoms counted in the original algorithm for the computation of line bundle cohomologies. The overall sign a rationom picks up under the bundle involution determines whether it contributes to the invariant or anti-invariant cohomology group, and non-trivial multiplicities apply canonically in this counting.
\end{description}

The simplicity of this (conjectured) algorithm to compute $\IZ_2$-equivariant cohomologies ultimately stems from the fact, that one can basically use the same involution mapping specified for the coordinates of the base toric space directly on the rationoms that represent the cohomology group---provided the used uplift of this mapping in the form of the equivariant structure has been specified appropriately. Quite recently the same conjecture was also posed and developed in the appendix of \cite{Cvetic:2010ky}, where the authors focus on the computation of the Lefschetz numbers---whose computation can become somewhat involved due to the equivariant Chern characters in \eqref{eq_holLefschetztheorem}---and also consider the standard examples $\IC\IP^1$, $\IC\IP^2$, $dP_1$ and $dP_3$ in detail. In the context of orientifolds we refer to their nice presentation of the $\IZ_2$-equivariant material.

\subsection{Invariants for finite group actions}
The mathematical background presented in section~\ref{sec_Z2invariants} can be applied to more involved finite group actions. However, some of the aspects loose their specific clarity that the special case of the two-element group $\IZ_2$ offers. Given a finite group
\beq
  G=\{ g_1,g_2,\dots,g_m\}
\eeq
of $m$ elements acting holomorphically on $X$, the relation \eqref{eq_Z2cosetformula} between the Euler characteristic of the orbifold space $X/G$ and the sum of the different Lefschetz numbers generalizes to
\beq
  \chi(X/G;V) = \frac{1}{|G|} \sum_{g\in G} \chi^g(X;V) = \sum_{i=0}^n(-1)^i h^i_{\rm inv}(X;V),
\eeq
where $|G|$ is the number $n$ of group elements. The index formula for the individual Lefschetz numbers \eqref{eq_holLefschetztheorem} remains unchanged, but has to be computed separately for the individual fixpoint sets of each group element. In the decomposition of the vector bundle $V$ into $g_*$-eigenbundles more general eigenvalues $\rho_k\in\IC$ can now arise. The computation of those eigenvalues rests on the group action on the conjugated normal bundle $\bar N_{X^g}$ of each component of the fixpoint set. Due to the decomposition
\beq\label{eq_tangentnormaldecomposition}
  TX|_{X^g} = TX^g \oplus N_{X^g}
\eeq
of the ambient space tangent bundle, the $g$-action on $N_{X^g}$ is given by a proper decomposition of the differential mapping
\beq\label{eq_differentialmapping}
  {\rm d} g_p : T_pX \longrightarrow T_{gp}X
\eeq
over a fixpoint $p=gp\in X^g$. In order to obtain the Lefschetz numbers, this then allows for the computation of the action's eigenvalues on $\bar N_{X^g}$ and the evaluation of the integral in \eqref{eq_holLefschetztheorem}.

\subsection{Some explicit examples for finite group equivariance}

\subsubsection*{Example: $\IC\IP^2/\IZ_3$}
As an example for  a generalization of the $\IZ_2$-conjecture posed on page~\pageref{Z2conjection} and in \cite{Cvetic:2010ky}, we consider the line bundle sheaf cohomology over the orbifold space $\IC\IP^2/\IZ_3$. Here the group action of $\IZ_3=\{e,g_1,g_2\}$ on $\IC\IP^2$ is defined by the generator
\beq\label{eq_CP2origInvolution}
  g_1:(u_1,u_2,u_3)\mapsto (\alpha u_1, \alpha^2 u_2, u_3) \qquad \text{for $\alpha\ce\sqrt[3]{1}={\rm e}^{\frac{2\pi{\rm i}}{3}}$}.
\eeq
Due to the projective equivalences between the homogeneous coordinates $u_i$ the mapping is equivalent to
\beq
  \bal
	  g'_1:(u_1,u_2,u_3) &\mapsto (u_1,\alpha u_2,\alpha^2 u_3) \\
		g''_1:(u_1,u_2,u_3)&\mapsto (\alpha^2 u_1, u_2, \alpha u_3),
	\eal
\eeq
i.e.~$g_1\sim g'_1\sim g''_1$ describe the same involution on the base space. Considering the Stanley-Reisner ideal ${\rm SR}(\IC\IP^2)=\langle u_1 u_2 u_3 \rangle$ this action therefore has three fixpoints
\beq
  P_1 = (0,0,1), \qquad P'_1=(1,0,0),\qquad P''_1=(0,1,0)
\eeq
in $\IC\IP^2$. The second group element's involution is given by the square
\beq
  \bal
    g_2 \ce g_1^2 : (u_1,u_2,u_3) &\mapsto (\alpha^2 u_1, \alpha u_2, u_3) \\
		g'_2: (u_1,u_2,u_3) &\mapsto (u_1,\alpha^2 u_2, \alpha u_3) \\
		g''_2:(u_1,u_2,u_3) &\mapsto (\alpha u_1, u_2, \alpha^2 u_3)
	\eal
\eeq
leading to the same three fixpoints $P_2 =P_1$, $P'_2=P'_1$ and $P''_2=P''_1$. Since for both non-trivial group elements the fixpoint sets consist of three components of maximal codimension --- isolated fixpoints --- this example is particularly simple.

In order to determine the (conjugated) normal bundle's eigenspace decomposition under the induced $\IZ_3$-action, we utilize that for fixpoints the general split \eqref{eq_tangentnormaldecomposition} leads to the direct identification $(N_{X^g})_p \cong T_pX$, i.e.~it suffices to compute the eigenvalues of the differentials \eqref{eq_differentialmapping} at the fixpoints. For the first fixpoint $P_1\in U_3=\{u_3\not=0\}\subset\IC\IP^2$ we use the local chart given by
\beq\label{eq_CP2localCharts}
  \bal
    \phi_3 : {}&{} U_3 \stackrel{\cong}{\longrightarrow} \IC^2 \\
	  &{} (u_1,u_2,u_3) \mapsto_\Big. \smash{\left(\frac{u_1}{u_3}, \frac{u_2}{u_3}\right)}.
	\eal
\eeq
The involution mapping $g_1$ within this chart then takes the form
\beq
  \bal
    f_1^3\ce \phi_3\circ g_1\circ \phi_3^{-1} : {}&{} \IC^2 \longrightarrow \IC^2 \\
		&{} (x,y) \mapsto (\alpha x, \alpha^2 y),
	\eal
\eeq
and the differential mapping at $\phi_3(P_1)=(0,0)\in\IC^2$ is then easily computed to
\beq
  {\rm d}(f_1^3)_{P_1} = \left(
	\begin{array}{cc}
	  \frac{\partial f^3_{1,x}}{\partial x}_\big.   &   \frac{\partial f^3_{1,x}}{\partial y}_\big. \\
		\frac{\partial f^3_{1,y}}{\partial x}         &   \frac{\partial f^3_{1,y}}{\partial y}
	\end{array}
	\right)_{\!\!\! P_1}
	= \ttmat{\alpha}{0}{0}{\alpha^2}
\eeq
Via $\det \big( {\rm d}(f_1^3)_{P_1} - \lambda \IMAT \big) = (\alpha-\lambda)(\alpha^2-\lambda)=0$ this leads to the eigenvalues $\lambda_1=\alpha$ and $\lambda_2=\alpha^2$, such that the action on the 2d${}_\IC$ conjugated normal bundle induces the split into $\IZ_3$-irreducible representations
\beq
  \bar N_{P_1} \cong \bar N_{P_1}^{\bar\alpha} \oplus \bar N_{P_1}^{\bar\alpha^2} \cong \bar N_{P_1}^\alpha \oplus \bar N_{P_1}^{\alpha^2}
\eeq
on the fixpoint $P_1$. The analogous computation yields the same result for all three fixpoints of both $g$ and $g^2$. Since $\dim X^g = \dim X^{g^2} = 0$ the expansion of the equivariant Chern character reduces to
\beq
  \bal
    {\rm ch}_g(&\Lambda_{-1} \bar N_{P_1}) = {\rm ch}_g(\cO - \bar N_{P_1} + \Lambda^2\bar N_{P_1}) \\
	  &= \dim\cO - (\alpha\dim\bar N_{P_1}^\alpha + \alpha^2\dim\bar N_{P_1}^{\alpha^2}) + \alpha\cdot\alpha^2 \dim \Lambda^2\bar N_{P_1} \\
	  &= 1-(\alpha+\alpha^2) + \alpha^3 = 1-(-1)+1 = 3.
	\eal
\eeq
Using ${\rm Td}(X^g)=1$ and ${\rm ch}_g(L)=\varrho_g(L;P)\in\IC^\times$ for each fixpoint component, it follows
\beq
  \bal
    \chi^g(\IC\IP^2;L) &= \left[\int_{P_1} + \int_{P'_1} + \int_{P''_1} \right] {\rm ch}_g(L)\frac{{\rm Td}(X^g)}{{\rm ch}_g(\Lambda_{-1}\bar N_{X^g})} \\
	  &= \frac{\varrho_g(L;P_1) + \varrho_g(L;P'_1) + \varrho_g(L;P''_1)}{3}, \\
    \chi^{g^2}(\IC\IP^2;L) &= \frac{\varrho_{g^2}(L;P_1) + \varrho_{g^2}(L;P'_1) + \varrho_{g^2}(L;P''_1)}{3}. \\
	\eal
\eeq

It remains to compute the eigenvalues $\varrho_g(L;P)$ that originate in the equivariant Chern character ${\rm ch}_g(\cO(k))$ of the line bundle, i.e.~we need to determine the irreducible representation of $\cO(k)$ under the $\IZ_3$-action. Using the so-called process of homogenization (see section~5.4 of \cite{Cox:ToricVarieties}) the divisor $D=kH$ that defines the bundle $\cO(D)$ can be represented by a monomial
\beq
  Q_{\IC\IP^2}(kH) = u_1^a u_2^b u_3^{k-a-b},
\eeq
where $a,b,k\in\IZ$. Whereas the strict definition of those monomials utilizes an inner product between certain lattice points related to the fan of $\IC\IP^2$ and the lattice points of the divisor $D$, the above form of such monomials can be easily read of from the GLSM charges, see \eqref{eq_dP1bdlMonomial} and \eqref{eq_dP3bdlMonomial} in the later examples. One can interpret the space of local sections of $\cO(kH)$ as generated by monomials of the form $u_1^a u_2^b u_3^c$ where $a+b+c=k$, i.e.~we can effectively use the monomial as a representation of the bundle. This representation bears a striking resemblance to our rationoms, cf.~\eqref{eq_CP3bundleMonomials}. The idea is then to apply the different (equivalent) base involutions $g_1,g'_1,g''_1$ associated to the fixpoints $P_1,P'_1,P''_1$ on this monomial and determine the value picked up relative to the involution that we choose for the equivariant structure, i.e.~the involution $g_1$ in this example. The choice of the equivariant structure for the bundle is therefore reflected in the bundle representation eigenvalues $\varrho_g(L;P)$. For our example we therefore have
\beq
  \xymatrix{                & \big.\smash{\overbrace{\alpha^{a+2b} Q}^{\mclap{\text{value of equivariant structure}}}} & \leadsto\quad \varrho_g(\cO(k);P_1)=1 \\
	Q=u_1^a u_2^b u_3^{k-a-b} \ar@(u,r)[ur]^{g_1\qquad} \ar@(d,r)[dr]_{g''_1\qquad} \ar[r]^{\quad g'_1}
	                          & \alpha^{-k}\alpha^{a+2b} Q & \leadsto\quad \varrho_g(\cO(k);P'_1)=\alpha^{-k} \\
	                          & \alpha^k \alpha^{a+2b} Q   & \leadsto\quad \varrho_g(\cO(k);P''_1)=\alpha^k}^\Big.
\eeq
and an analogous result for $g_2$, leading to the final expressions
\beq
  \chi^g\big(\IC\IP^2;\cO(k)\big) =
	\chi^{g^2}\big(\IC\IP^2;\cO(k)\big) =
	\frac{1+\alpha^k+\alpha^{-k}}{3} =
	\begin{cases} 1 & k\in 3\IZ \\ 0 & \text{otherwise} \end{cases}\; .
\eeq
Together with the ordinary Euler characteristic of $\cO(k)$ on $\IC\IP^2$
\beq
  \chi(\IC\IP^2;\cO(k)) = 1+\frac{k(k+3)}{2}
\eeq
we therefore obtain the orientifold Euler characteristic
\beq\label{eq_CP2Z3exLefschetzNumber}
  \bal
    \chi\big(\IC\IP^2/\IZ_3;\cO(k)\big) &= \frac{\chi+\chi^g+\chi^{g^2}}{3} \\
	  &= \frac{6+3k(k+3) + 4(1+\alpha^k+\alpha^{-k})}{18},
	\eal
\eeq
which completes the computation on the well-established and proven mathematical side.

The idea is now to simply apply the involution mapping to the rationoms of our counting algorithm and count the remaining invariant rationoms. Recall from section~\ref{sec_Z2algconjecture} that this already implies a choice of the equivariant $\IZ_3$-structure on the bundle, where we will use the non-primed involution mapping $g_1$. Consider for example the bundle $\cO(-6)$ on $\IC\IP^2$. From \eqref{eq_CP2Z3exLefschetzNumber} we expect to find $\chi(\IC\IP^2/\IZ_3;\cO(-6))=4$. The relevant algorithm rationoms and their respective phases picked up from the involution are
\beq\label{eq_CP2algorithmCounting}
  \underbrace{
  \bal
	  \underbrace{\frac{1}{u_1^4u_2u_3}}_{g_1\to 1}, \quad
		\underbrace{\frac{1}{u_1u_2^4u_3}}_{g_1\to 1}, \quad
		\underbrace{\frac{1}{u_1u_2u_3^4}}_{g_1\to 1}, \quad
		\underbrace{\frac{1}{u_1^3u_2^2u_3}}_{g_1\to\alpha}, \quad
		\underbrace{\frac{1}{u_1^3u_2u_3^2}}_{g_1\to\alpha^2}, \\
		\underbrace{\frac{1}{u_1^2u_2^3u_3}}_{g_1\to\alpha^2}, \quad
		\underbrace{\frac{1}{u_1u_2^3u_3^2}}_{g_1\to\alpha}, \quad
		\underbrace{\frac{1}{u_1^2u_2u_3^3}}_{g_1\to\alpha}, \quad
		\underbrace{\frac{1}{u_1u_2^2u_3^3}}_{g_1\to\alpha^2}, \quad
		\underbrace{\frac{1}{u_1^2u_2^2u_3^2}}_{g_1\to 1},
	\eal}_{\displaystyle 	h^2(\IC\IP^2;\cO(-6))=(4_{\text{inv}},3_\alpha,3_{\alpha^2})}
\eeq
yielding $h^\bullet_{\text{inv}}(\IC\IP^2;\cO(-6))=(0,0,4)$ and therefore the expected result for the Euler characteristic of the orbifold space $\IC\IP^2/\IZ_3$. Note that due to $g_2 = g_1^2$ one only has to evaluate the effect of the generators to identify the invariant rationoms. This agreement has been checked for a wide range of bundles $\cO(k)$ on $\IC\IP^2/\IZ_3$. 

\subsubsection*{Example: $dP_1/\IZ_3$}
\begin{table}[t]
  \centering
  \begin{tabular}{r@{\,$=$\,(\,}r@{,\;\;}r@{\,)\;\;}|c|cc|c}
    \multicolumn{3}{c|}{vertices of the} & coords & \multicolumn{2}{c|}{GLSM charges} & {divisor class}${}^\big.$ \\
    \multicolumn{3}{c|}{polyhedron / fan} & & $Q^m$ & $Q^n$ & \\
    \hline\hline
    $\nu_1$ & $-1$ & $-1$ & $u_1$ & 1 & 0 & $H$\uspc \\
    $\nu_2$ &  1   &  0   & $u_2$ & 1 & 0 & $H$  \\
    $\nu_3$ &  0   &  1   & $u_3$ & 1 & 1 & $H+X$ \\
    $\nu_4$ &  0   & $-1$ & $u_4$ & 0 & 1 & $X$
  \end{tabular}
  \\[5mm] intersection form: ${}\quad HX - X^2$
  \\[3mm] ${\rm SR}(dP_1) = \langle x_1 x_2 ,\; x_3 x_4 \rangle$
  \caption{\small Toric data for the del Pezzo-1 surface}
  \label{tab:dPoneSurface}
\end{table}

Next we consider a blowup of $\IC\IP^2$, i.e.~the del~Pezzo-1 surface. The involution \eqref{eq_CP2origInvolution} basically remains unchanged, acting now on the four homogeneous coordinates of $dP_1$ as
\beq\label{eq_dP1origInvolution}
  g:(u_1,u_2,u_3,u_4)\mapsto (\alpha u_1, \alpha^2 u_2, u_3, u_4) \qquad \text{for $\alpha\ce\sqrt[3]{1}={\rm e}^{\frac{2\pi{\rm i}}{3}}$}.
\eeq
Following from the projective equivalences listed in table~\ref{tab:dPoneSurface}, we can identify the four fixpoints of the action:
\beq
  P_1 = (1,0,0,1), \quad
	P_2 = (0,1,0,1), \quad
	P_3 = (0,1,1,0), \quad
	P_4 = (1,0,1,0).
\eeq
By using local charts around those fixpoints like in \eqref{eq_CP2localCharts}, the tangent space mapping eigenvalues reveal the following representations for the conjugated normal bundles:
\beq
  \bar N_{P_1} = \bar N_{P_1}^\alpha \oplus \bar N_{P_1}^{\alpha^2}, \quad
	\bar N_{P_2} = \bar N_{P_2}^\alpha \oplus \bar N_{P_2}^{\alpha^2}, \quad
	\bar N_{P_3} = (\bar N_{P_3}^\alpha)^2, \quad
	\bar N_{P_4} = (\bar N_{P_4}^{\alpha^2})^2.
\eeq
Compared to the three $\IC\IP^2$ fixpoints of the analogous $\IZ_3$-action, whose representations were all of the type $\bar N_P^\alpha\oplus \bar N_P^{\alpha^2}$, the additional blowup of $dP_1$ seems to split up the contribution of one of three $\IC\IP^2$ fixpoints. This can be seen by
\beq
  \bal
	  {\rm ch}_g\big((\bar N_P^\alpha)^2\big)\big|_P &{}= 1-\alpha\cdot 2 + \alpha^2 = (1-\alpha)^2 \\
		{\rm ch}_g\big((\bar N_P^{\alpha^2})^2\big)\big|_P & {}= 1-\alpha^2\cdot 2+\alpha^4 = (1-\alpha^2)^2
	\eal
\eeq
and noting that the sum of both these contributions adds up to
\beq
  \frac{1}{(1-\alpha)^2} + \frac{1}{(1-\alpha^2)^2} = \frac{1}{3},
\eeq
i.e.~precisely the contribution that each $\IC\IP^2$ fixpoint added to the Lefschetz numbers in the previous example. The local sections of the bundle $\cO(m,n)$ over $dP_1$ can be represented by monomials of the form
\beq\label{eq_dP1bdlMonomial}
  Q_{dP_1}(mH+nX) = u_1^a u_2^b u_3^{m-a-b} u_4^{n-m+a+b},
\eeq
and relative to the involution $g$ from \eqref{eq_dP1origInvolution} (that we choose for the equivariant structure) this gives the relative signs, fixpoints and normal bundle representation splittings
\beq\label{eq_dP1fixpointData}
  \begin{array}{lclcccl}
	  P_1 = (1,0,0,1)\Big.  && \bar N_{P_1} = \bar N_{P_1}^\alpha \oplus \bar N_{P_1}^{\alpha^2} && \frac{1}{3}              && \alpha^{-m} \\
	  P_2 = (0,1,0,1)\Big.  && \bar N_{P_2} = \bar N_{P_2}^\alpha \oplus \bar N_{P_2}^{\alpha^2} && \frac{1}{3}              && \alpha^{m} \\
	  P_3 = (0,1,1,0)\Big.  && \bar N_{P_3} = (\bar N_{P_3}^\alpha)^2                                && \frac{1}{(1-\alpha)^2}   && \alpha^{m-n} \\
	  P_4 = (1,0,1,0)\Big.  && \bar N_{P_4} = (\bar N_{P_4}^{\alpha^2})^2                            && \frac{1}{(1-\alpha^2)^2} && \alpha^{-(m-n)}
	\end{array}
\eeq
>From the standard Riemann-Roch-Hirzebruch formula \eqref{eq_riemannrochhirzebruch} one can compute the ordinary Euler characteristic
\beq
  \chi(dP_1;\cO(m,n)) = 1 + m + mn + \frac{1}{2}n(1-n)
\eeq
and from the fixpoint data listed in \eqref{eq_dP1fixpointData} the Lefschetz number of the generator $g$ can be evaluated as
\beq\label{eq_dP1firstLefschetzNum}
  \chi^g(dP_1;\cO(m,n)) = \frac{\alpha^{-m} + \alpha^m}{3} + \frac{\alpha^{m-n}}{(1-\alpha)^2} + \frac{\alpha^{-(m-n)}}{(1-\alpha^2)^2}.
\eeq
Note that this Lefschetz number is not an integer for generic values of $m,n\in\IZ$. However, since for the $\IZ_3$ group the direct identification \eqref{eq_normalsplitting} of the single Lefschetz number $\chi^g$ with dimensions of cohomology groups is no longer given, this does not pose a problem. One can show that the Lefschetz number for the second non-unit group element $g^2\in\IZ_3$ can be obtained from replacing $\alpha \to \alpha^2$ in formula \eqref{eq_dP1firstLefschetzNum}. Ultimately, we therefore arrive at the Euler characteristic
\beq
  \bal
    &\chi(dP_1/\IZ_3;\cO(m,n)) = \frac{\chi + \chi^g + \chi^{g^2}}{3} \\
		&{}= \frac{1}{3}\Bigg[ 1 + m + mn + \frac{n(1-n)}{2}+\frac{2(\alpha^{-m}+\alpha^m)}{3} + \frac{2\alpha^{m-n}}{(1-\alpha)^2} + \frac{2\alpha^{-(m-n)}}{(1-\alpha^2)^2} \Bigg]
	\eal
\eeq
for the orbifold space obtained from the $\IZ_3$-action on the single blowup of $\IC\IP^2$. Turning to the counting of $g$-invariant rationoms from our algorithm as in \eqref{eq_CP2algorithmCounting}, we find once again perfect agreement with the Euler characteristic derived from the Lefschetz theorem above.

\subsubsection*{Example: $dP_3/\IZ_3$}
\begin{table}[t]
  \begin{center}
    \begin{tabular}{r@{\,$=$\,(\,}r@{,\;\;}r@{\,)\;\;}|c|cccc|c} 
      \multicolumn{3}{c|}{vertices of the} & coords & \multicolumn{4}{c|}{GLSM charges} & {divisor class}${}^\big.$ \\
      \multicolumn{3}{c|}{polyhedron / fan}      &        & $Q^m$ & $Q^n$ & $Q^p$ & $Q^q$ & \\ 
			\hline\hline
      $\nu_1$ & $-1$ & $-1$ & $u_1$ & 1 & 0 & 0 & 1 & $H+Z$\uspc \\
      $\nu_2$ &   1  &   0  & $u_2$ & 1 & 0 & 1 & 0 & $H+Y$ \\
      $\nu_3$ &   0  &   1  & $u_3$ & 1 & 1 & 0 & 0 & $H+X$ \\
      $\nu_4$ &   0  & $-1$ & $u_4$ & 0 & 1 & 0 & 0 & $X$ \\
      $\nu_5$ & $-1$ &   0  & $u_5$ & 0 & 0 & 1 & 0 & $Y$ \\
			$\nu_6$ &   1  &   1  & $u_6$ & 0 & 0 & 0 & 1 & $Z$ 
    \end{tabular}
    \\[5mm] intersection form: ${}\quad HX + HY + HZ - 2H^2 - X^2 - Y^2 - Z^2$
		\\[3mm] ${\rm SR}(dP_3) = \langle x_1 x_2,\; x_1 x_3, \; x_1 x_6, \; x_2 x_3, \; x_2 x_5, \; x_3 x_4, \; x_4 x_5, \; x_4 x_6, \; x_5 x_6 \rangle$
    \vspace{-3mm}
  \end{center}
  \caption{\small Toric data for the del Pezzo-3 surface.}
  \label{tbl:dPthreeSurface}
\end{table}

The natural extension to the previous example is to blowup the $dP_1$ twice further, giving us the $dP_3$ surface with the toric data in table~\ref{tbl:dPthreeSurface}. Once again we employ the same extension of the action \eqref{eq_CP2origInvolution} on the base. However, in order to simplify the subsequent computation, this time we choose a different equivariant structure, which is induced via
\beq\label{eq_dP3origInvolution}
  \bal
    g:(u_1,\dots,u_6)\mapsto{}&{}(u_1,\alpha u_2,u_3,\alpha u_4,u_5,u_6) \\
		\sim {}&{} (\alpha u_1, \alpha^2 u_3, u_3, u_4, u_5, u_6).
	\eal
	\qquad \text{for $\alpha\ce\sqrt[3]{1}={\rm e}^{\frac{2\pi{\rm i}}{3}}$}.
\eeq
The $\IZ_3$-action on $dP_3$ reveals six fixpoints, which can be related to the ``splitting'' of each of the three $\frac{1}{3}$-fixpoint contributions from the original $\IC\IP^2$ computation. Computing the induced normal bundle representation and relative signs via
\beq\label{eq_dP3bdlMonomial}
  Q_{dP_3}(mH+nX+pY+qZ)=u_1^a u_2^b u_3^{m-a-b} u_4^{n-m+a+b} u_5^{p-b} u_6^{q-a}
\eeq
is completely analogous --- albeit quite laborious --- to the previous cases and yields the following fixpoint data:
\beq
  \begin{array}{lclcccl}
	  P_1 = (1,0,1,0,1,1)\Big.  && \bar N_{P_1} = (\bar N_{P_1}^{\alpha^2})^2 && \frac{1}{(1-\alpha^2)^2} && 1 \\
		P_2 = (0,1,1,1,0,1)\Big.  && \bar N_{P_2} = (\bar N_{P_2}^{\alpha^2})^2 && \frac{1}{(1-\alpha^2)^2} && \alpha^{m-n+p} \\
		P_3 = (1,1,0,1,1,0)\Big.  && \bar N_{P_3} = (\bar N_{P_3}^{\alpha^2})^2 && \frac{1}{(1-\alpha^2)^2} && \alpha^{q-m-n} \\
		P_4 = (0,1,1,0,1,1)\Big.  && \bar N_{P_4} = (\bar N_{P_4}^{\alpha})^2   && \frac{1}{(1-\alpha)^2}   && \alpha^{n-m} \\
		P_5 = (1,0,1,1,1,0)\Big.  && \bar N_{P_5} = (\bar N_{P_5}^{\alpha})^2   && \frac{1}{(1-\alpha)^2}   && \alpha^{m-n-q} \\
		P_6 = (1,1,0,1,0,1)\Big.  && \bar N_{P_6} = (\bar N_{P_6}^{\alpha})^2   && \frac{1}{(1-\alpha)^2}   && \alpha^{-n-p} \\
	\end{array}\; .
\eeq
Employing once again the well-known Riemann-Roch-Hirzebruch formula, the Euler characteristic of $dP_3$ turns out to be
\beq\label{eq_dP3fixpointData}
  \bal
    \chi(dP_3,\cO(m,n,p,q)) = {}&{} 1 - m^2 + mn + mp + mq \\ &{} + \frac{n(1-n) + p(1-p) + q(1-q)}{2}
	\eal
\eeq
and from the fixpoint data in \eqref{eq_dP3fixpointData} we can compute the Lefschetz number
\beq\label{eq_dP3firstLefschetzNum}
  \bal
		\chi^g(dP_3;\cO(m,n,p,q)) 
		={}&{} \frac{1+\alpha^{m-n+p}+\alpha^{q-m-n}}{(1-\alpha^2)^2} \\
		{}&{}+ \frac{\alpha^{n-m}+\alpha^{m-n-q}+\alpha^{-n-p}}{(1-\alpha)^2} .
	\eal
\eeq
Again, this number will not be an integer for a generic choice of the bundle divisor $D=mH+nX+pY+qZ$. The second Lefschetz number $\chi^{g^2}$ can be obtained by replacing $\alpha\to\alpha^2$ in formula \eqref{eq_dP3firstLefschetzNum}, such that the average of the three terms gives us the Euler characteristic of the orbifold space $dP_3/\IZ_3$:
\beq\label{eq_dP3Z3Lefschetz}
  \bal
    \chi(dP_3/\IZ_3;{}&{}\cO(m,n,p,q)) = 1 - m^2 + mn + mp + mq \\ 
	  &{} + \frac{n(1-n) + p(1-p) + q(1-q)}{2} + \frac{1}{3} \\
	  &{}+\frac{\alpha^{m-n} + \alpha^{n+p} + \alpha^{m-n+p} + \alpha^{-m+n+q} + \alpha^{-m-n+q}}{(1-\alpha^2)^2} \\
	  &{}+\frac{\alpha^{-m+n} + \alpha^{-n-p} + \alpha^{-m+n-p} + \alpha^{m-n-q} + \alpha^{m+n-q}}{(1-\alpha)^2}.
	\eal
\eeq
By comparison to the rationom counting of our algorithm, we find once again perfect agreement. In addition to simply providing a more complicated example, the $dP_3$ rationom counting also involves non-trivial multiplicity factors~2. Consider for example the line bundle $\cO(-5,-1,-1,-1)$, which has six monomials each contributing with multiplicity factor~2. Applying the involution \eqref{eq_dP3origInvolution} (that was chosen for the equivariant structure and therefore directly acts on the rationoms) we observe the following:
\beq\label{eq_dP3algorithmCounting}
  \underbrace{
	 2\times\,\,\Bigg( \, \underbrace{\frac{u_4^2}{u_1 u_2 u_3^3}}_{g\to \alpha}, \quad
		\underbrace{\frac{u_4 u_5}{u_1 u_2^2 u_3^2}}_{g\to \alpha^2}, \quad
		\underbrace{\frac{u_5^2}{u_1 u_2^3 u_3}}_{g\to 1}, \quad
		\underbrace{\frac{u_4 u_6}{u_1^2 u_2 u_3^2}}_{g\to 1}, \quad
		\underbrace{\frac{u_5 u_6}{u_1^2 u_2^2 u_3}}_{g\to\alpha}, \quad
		\underbrace{\frac{u_6^2}{u_1^3 u_2 u_3}}_{g\to\alpha^2} \, \Bigg)
	}_{\displaystyle 	h^1(dP_3;\cO(-5,-1,-1,-1))=(4_{\text{inv}},4_\alpha,4_{\alpha^2})}
\eeq
Plugging the bundle charges into \eqref{eq_dP3Z3Lefschetz} yields 
\beq
  \chi(dP_3/\IZ_3;\cO(-5,-1,-1,-1))=-4,
\eeq
once again in agreement with the result obtained from the counting of invariant rationoms. Similar to the observation made for $\IZ_2$-equivariant situations we therefore find the same ``canonical'' behavior of such rationoms, i.e.~an invariant rationom with multiplicity~2 simply contributes twice to the counting of invariant rationoms.

\subsection{Generalized equivariant algorithm conjecture}\label{sec_FiniteGroupConjecture}
The steps involved in the computation of the Lefschetz character in the $\IC\IP^2$ example are completely analogous on $\IC\IP^{n-1}$ with the group $\IZ_n$ and the base space generator involution 
\beq
  g:u_i \mapsto \alpha^i u_i \qquad \text{with }\alpha\ce\sqrt[p]{1} = {\rm e}^{\frac{2\pi{\rm i}}{p}}.
\eeq
We have successfully checked this for various values of $n$ and bundles $\cO(k)$. Together with the empirical evidence gathered from the presented and various other examples, we therefore arrive at the following hypothesis which generalizes the conjecture from section~\ref{sec_FiniteGroupConjecture}:

\begin{description}
  \item[Conjecture for {\itshape\bfseries G}-equivariant cohomology of finite groups:]\label{FiniteGconjecture} Given a toric space $X$, generator involutions on the base $\sigma_1,\dots,\sigma_r:X\fto X$ as well as an equivariant structure, the lifted involution mapping can be directly applied to the rationoms counted in the original algorithm for the computation of line bundle cohomologies. The rationoms entirely invariant under all generator mappings contribute to the invariant cohomology (i.e.~the cohomology on the orbifold  space $X/G$) and non-trivial multiplicities apply canonically in this counting.
\end{description}

At this point we would like to emphasize the tremendous computational power of this conjecture. Already for the $dP_3$ example --- still a rather simple surface --- we see that the computations necessary to just determine the Euler characteristic (much less than the cohomology groups themselves) from the established mathematics is quite enduring, whereas this computation for reasonably low values of the bundle charges can be done via pen and paper in a couple of minutes. As before the efficiency and ease-of-usage this conjecture provides ultimately rests on the direct applicability of the involution mappings defined on the coordinates of the base on the rationoms representing the cohomology.

\section{Connection to combinatorial toric geometry}\label{sec_genbatyrev}
So far, in order to obtain information about the cohomologies of various bundles, we have always been working with exact sequences. It is well known that there also exist a combinatorial approach to calculate such quantities via certain lattice polytopes that contain the toric data. The first description of the Hodge numbers of a Calabi-Yau hypersurface in this fashion \cite{Batyrev} was followed by the generalization to complete intersections in higher-dimensional ambient spaces \cite{BatBor}. Recently, there were also attempts to calculate bundle deformations of the tangent bundle of a Calabi-Yau 3-fold in such a way \cite{KreuzerMelnikov,02MirrorMap}. In this chapter we want to show how it is possible to relate the ingredients of such calculations to cohomology classes of line bundles of the corresponding ambient space which may allow a deeper insight to the combinatorial formulas.

\subsection{Lattice polytopes and Calabi-Yau hypersurfaces}\label{subsection_Batyrev hypersurface}
As mentioned earlier, the geometry of a toric variety can be described by its fan which itself is defined as a triangulation of a given reflexive polytope. A lattice polytope is called reflexive, if its polar polytope is a lattice polytope, as well. Let $\Delta^\circ$ be such a reflexive polytope, then its polar polytope $\Delta$ is defined as
\beq
  \Delta \ce \left\{ m \in \mathbb{Z}^d : \left<n,m\right> \geq -1~\forall n\in \Delta^\circ\right\}.
\eeq

While the vertices of the reflexive polytope $\Delta^\circ$ of a given toric variety $X$ represent the homogeneous coordinates as well as the equivalence relations between them, the lattice points of the polar polytope $\Delta$ represent a Calabi-Yau hypersurface $M$ in $X$. This hypersurface is defined by an equation $F=0$, where $F$ is a sum of monomials induced from the lattice points $m$ of $\Delta$. The term corresponding to a fixed $m$ is then proportional to
\beq\label{eq:hypersurface monomials}
  \prod_{\rho\in \Delta^\circ}z_\rho^{\left<m,\rho\right>+1}.
\eeq
The polytope $\Delta$ is also called the Newton polytope of $M$. On the other hand we can consider the vertices of $\Delta$ to be the defining data of some other toric variety and its polar polytope $\Delta^\circ$ to be the Calabi-Yau hypersurface of $\Delta$. In this way we are able to relate two apparently completely different Calabi-Yau varieties with each other. For such Calabi-Yau hypersurfaces this relation is called mirror symmetry, see \cite{HoriEA:MirrorSymmetry} for an exhaustive treatment. Since $F$ is made out of terms from \eqref{eq:hypersurface monomials} with different coefficients and due to the fact that $F$ describes the Calabi-Yau manifold, it is clear that the complex structure deformations are very much depending on $\Delta$. On the other hand, vertices in the polytope $\Delta^\circ$ correspond to homogeneous coordinates $x_k$ and can therefore define the usual coordinate-associated divisors
\beq
  D_k:=\left\{x_k=0\right\}
\eeq
on the Calabi-Yau. Hence it is reasonable to assume that $\Delta^\circ$ strongly influences the K\"ahler moduli. 

The mirror symmetry conjecture, as it follows trivially from conformal field theory, is geometrically highly non-trivial and states that for every Calabi-Yau threefold $M$ with Hodge numbers $(h_{21}(M),h_{11}(M))$ there exists a mirror dual $\hat M$ one with exchanged Hodge numbers $h_{21}(\hat M)=h_{11}(M)$ and   $h_{11}(\hat M)=h_{21}(M)$. Thus, mirror symmetry exchanges complex structure deformations of $M$ with the K\"ahler deformations of $\hat M$ and vice versa. Mirror symmetry for hypersurfacses in $d$-dimensional toric varieties has  the  precise mathematical meaning of a simple exchange of the role of the two polytopes $\Delta$ and $\Delta^\circ$. This is reflected in a nice way in the well-known combinatorial formula for the Hodge numbers of a Calabi-Yau hypersurface, first derived by \cite{Batyrev}:
\begin{align}
	\label{eq:Batyrev1}
	h^{1,1}(M)&=l(\Delta^\circ)-d-1-\sum_{\mclap{\textrm{dim}(y)=0}}l^\ast(y^\vee)+
	  \sum_{\mclap{\textrm{dim}(y)=1}} l^\ast(y)\cdot l^\ast(y^\vee),\\
	\label{eq:Batyrev2} 
	h^{d-2,1}(M)&=l(\Delta)-d-1-\sum_{\mclap{\textrm{dim}(y_\circ)=0}} l^\ast(y_\circ^\vee)+
	  \sum_{\mclap{\textrm{dim}(y_\circ)=1}} l^\ast(y_\circ)\cdot l(y_\circ^\vee).
\end{align}
Here $y$ and $y_\circ$ are faces of $\Delta$ and $\Delta^\circ$, respectively, $l$ is the number of points in a face and $l^\ast$ is the number of interior points of a face. The dual face of an $r$ dimensional face $y\subset\Delta$ is a $(d-r-1)$-dimensional face $y^\vee\subset\Delta^\circ$ defined by
\beq
  y^\vee \ce \left\{n\in\Delta^\circ:\left<n,m\right> =-1~\forall~m\in y \right\}.
\eeq
{}From the two equations \eqref{eq:Batyrev1} and \eqref{eq:Batyrev2} it is clear that the Calabi-Yau hypersurfaces $M_\Delta$ and $M_{\Delta^\circ}$ associated to the polytopes $\Delta$ and $\Delta^\circ$ have mirror symmetric Hodge numbers. The Batyrev formula \eqref{eq:Batyrev1} counts the K\"ahler deformations of a given Calabi-Yau variety while the complex structure deformations are counted by equation \eqref{eq:Batyrev2}. The first terms in those equations correspond to toric and polynomial deformations while the last term, where faces and dual faces get mixed up, corresponds to non-toric, i.e.~non-polynomial deformations, respectively.

\subsubsection*{The Batyrev formulas in terms of line bundle cohomology}
As we have seen explicitly in section~\ref{subsection:Hypersurface}, it is also possible to obtain the Hodge numbers by making use of the Euler sequence \eqref{eq_longexacttangent} and the Koszul complex \eqref{eq_koszulsequence}. Therefore the contributions to equations \eqref{eq:Batyrev1} and \eqref{eq:Batyrev2} have their origin in the cohomology of line bundles on the ambient space. The observation for three dimensional hypersurfaces  is that all contributions to the polynomial deformations in the Batyrev formula arise from global sections of  line bundles on the ambient space, namely from $h_X^0(\cdot)$ for some divisor. In contrast, the non-polynomial deformations of the complex structure, which can also be associated with the twisted sector of the corresponding Gepner model, arise as $h_X^1(\cdot)$ contributions. 

In summary, we have observed the following identification of the various combinatorial contributions in the Batyrev formula with line bundle cohomologies:
\beq
	\boxed{
	\bal
		\fhh^{2,1}_0(M)&=l(\Delta)-d-1-\sum_{\mclap{\textrm{dim}(y_\circ)=0}} l^\ast(y_\circ^\vee)\,, 	\\
		\fhh^{2,1}_1(M)&=\sum_{\mclap{\textrm{dim}(y_\circ)=1}} l^\ast(y_\circ)\cdot l(y_\circ^\vee)\,,
	\eal
	}
\eeq
where we used the notation $\fhh^{p,q}_i$ introduced in \eqref{eq_weirdcontributionnumbers}. For the hypersurface case we could not find any example where the second or any higher cohomology contributed to this particular Hodge number.

For a K3 surface embedded in $\mathbb P^4$ the situation is a little different. Since it has only dimension~2, the tangent bundle cohomology ends of course with $h^2_M(T_M)$. Usually the $h_X^0(\cdot)$ and $h_X^1(\cdot)$ contributions ``flow'' in the complex structure deformations and the higher ones like $h_X^{n-1}(\cdot)$ and  $h_X^n(\cdot)$ to the K\"ahler deformations. Since the K3 has only a single non-trivial Hodge number that is not on the edge of the Hodge diamond, both sides contribute to it and we can see a split of this Hodge number $h^{1,1}_\text{K3}=20$ into $19+1$. We also know that for this K3 there is a split of the 20~deformations into 19~algebraic ones which would correspond to those coming from the $h_X^0(\cdot)$ and 1~non-algebraic one associated to the $h_X^2(\cdot)$ contributions.

One can easily check this with the {\cohomCalg} Koszul extension \cite{cohomCalg:Implementation} by using the ``{\tt Verbose5}'' option and following the contributions through the long exact sequences.

\subsection{Cayley polytopes and CICYs}
The next step is to extend the above formula to a formula for a complete intersection Calabi-Yau (CICY) in a 5-dimensional toric variety. This was done by Batyrev and Borisov shortly after the hypersurface case \cite{BatBor}. We will now describe how to do that, following \cite{Novoseltsev}.

\subsubsection*{Nef partitions and their Cayley polytopes}
 As before we start with a reflexive polytope $\Delta^\circ$, which is this time describing a 5-dimensional toric variety $X$. 
Consider now a partition of all vertices in $\Delta^\circ$ into disjoint subsets $V_i$ and define the polytopes following from those vertices as
\beq
  \Delta^\circ_i \ce \text{Conv}(V_i,0)\quad\text{for $i = 1,\dots,l$},
\eeq
where Conv$(\cdot)$ denotes the convex hull of a vertex set. Such a partition is called a nef (numerically effective) partition, if the Minkowski sum of all $\Delta^\circ_i$ forms a reflexive polytope. As a reminder, the Minkowski sum of two vertex sets is defined by
\beq
  \Delta^\circ_1+\Delta^\circ_2 \ce \text{Conv}(\left\{n+n':n\in\Delta^\circ_1,~n'\in\Delta^\circ_2\right\})\,.
\eeq
It is easy to see that one can associate a complete intersection Calabi-Yau variety to such a partition. It is the intersection of $l$ hypersurfaces, where each hypersurface $S_i$ corresponds to a sum of divisors $D_{i,k}=\{x_{k} = 0\}$. Here each $x_k$ is the homogeneous coordinate that belongs to the vertex $n_k\in\Delta^\circ_i$:
\beq
  S_i=\sum_{k\in K} D_{i,k}\,, \text{ where } K=\left\{k\in\mathbb N \text{ such that } n_k\in \Delta^\circ_i\right\}\,.
\eeq

So one might assume that one treats those ``subpolytopes'' in the same fashion as in the hypersurface case, but this is not quite the right procedure. Instead of dealing with the polytope that describes the toric ambient space, it is necessary to construct a different kind of polytope that respects the nef partition explicitly. This polytope is called the Cayley polytope and is defined by
\beq\label{eq:Cayley polytope}
  P^\ast\ce \text{Conv}(\Delta^\circ_1 \times e_1,\dots,\Delta^\circ_l \times e_l).
\eeq
Here $e_1,\dots,e_l$ is the canonical basis of $\mathbb Z^l$ and hence the Cayley polytope is a $d+l$ dimensional polytope, requiring a somewhat more sophisticated treatment compared to the aforementioned cases. For instance, it is not possible to compute a well-defined polar polytope anymore, and we need a different way to find a dual polytope for the Cayley polytope. It goes as follows: We consider the cone that supports the Cayley polytope, called the Cayley cone $C^\ast$, and calculate its dual cone $C$ via
\beq
  C\ce\left\{n\in \mathbb R^d\times \mathbb R^l : \left<m,n\right> \geq 0,~\forall m\in C^\ast\right\}.
\eeq
One can see that the dual Cayley cone $C$ also supports a polytope called the dual Cayley polytope. If we take slices of this one by successively putting the $i$th coordinate in $\left\{x_{d+1},\dots,x_{d+i},\dots,x_{d+l}\right\}$ to one and all the others in this set to zero for $i=1,\dots,l$, we get polytopes $\nabla_i$ whose Minkowski sum corresponds to the polar polytope $\Delta$ of our original polytope $\Delta^\circ$, namely
\beq
  \Delta=\nabla_1+\dots+\nabla_l.
\eeq
On the other hand, the convex hull $\nabla=\text{Conv}(\left\{\nabla_i,~i=1,\dots,l\right\})$ also represents a reflexive polytope that is polar to the Minkowski sum of the dual nef partition, i.e.~$\nabla^\circ = \Delta^\circ_1+\ldots+\Delta^\circ_l$ which gives rise to a complete intersection Calabi-Yau in the toric variety corresponding to this polytope. In summary, we therefore have the following relation between $\Delta^\circ$ and $\nabla$:
\beq\label{eq_dualize Cayley}
  \bal
    & \Delta^\circ  = \text{Conv}(\left\{\Delta^\circ_i,~i=1,\dots,l\right\}), && \nabla=\text{Conv}(\left\{\nabla_i,~i=1,\dots,l\right\}), \\
    & \Delta  = \nabla_1+\ldots+\nabla_l, && \nabla^\circ = \Delta^\circ_1+\ldots+\Delta^\circ_l\,,\\
    & P^\ast_{\Delta^\circ} = P_\nabla\,, && P^\ast_{\nabla} = P_{\Delta^\circ}\,.
	\eal
\eeq

So we need two different polytopes in order to find the two dual complete intersection Calabi-Yau manifolds. One of them is the complete intersection of hypersurfaces corresponding to vertex sets $\Delta_i^\circ$ that intersect in the toric variety built from $\Delta^\circ$, the other one is a complete intersection of hypersurfaces corresponding to vertex sets $\nabla_i$ that intersect in the toric variety built from $\nabla$. Clearly for the hypersurface case we find that $\nabla=\Delta$ and $\nabla^\circ=\Delta^\circ$ which reduces everything to the setting of section \ref{subsection_Batyrev hypersurface}.

\subsubsection*{Example: $\IP^5_{112233}$}
\begin{table}[ht]
  \centering
  \subfloat[Lattice polytope of $\IP^5_{112233}$]{
		\centering
		\begin{tabular}{r@{\,$=$\,(\,}r@{,\;\;}r@{,\;\;}r@{,\;\;}r@{,\;\;}r@{\,)\;\;}}
			\multicolumn{6}{c}{Vertices of $\Delta^\circ$}\\
			\hline\hline
			$\nu_1{}^\big.$ & $-1$ & $-2$ & $-2$ & $-3$ & $-3$\\
			$\nu_2$ & $ 1$ & $ 0$ & $ 0$ & $ 0$ & $ 0$\\
			$\nu_3$ & $ 0$ & $ 1$ & $ 0$ & $ 0$ & $ 0$\\
			$\nu_4$ & $ 0$ & $ 0$ & $ 1$ & $ 0$ & $ 0$\\
			$\nu_5$ & $ 0$ & $ 0$ & $ 0$ & $ 1$ & $ 0$\\
			$\nu_6$ & $ 0$ & $ 0$ & $ 0$ & $ 0$ & $ 1$
		\end{tabular}
		\label{tab_P112233}
	}
	\subfloat[Polar polytope]{
		\centering
		\begin{tabular}{r@{\,$=$\,(\,}r@{,\;\;}r@{,\;\;}r@{,\;\;}r@{,\;\;}r@{\,)\;\;}}
			\multicolumn{6}{c}{Vertices of $\Delta$}\\ 
			\hline\hline
			$\mu_1{}^\big.$ & $-1$ & $-1$ & $-1$ & $-1$ & $-1$\\
			$\mu_2$ & $ 3$ & $-1$ & $-1$ & $-1$ & $-1$\\
			$\mu_3$ & $-1$ & $ 3$ & $-1$ & $-1$ & $-1$\\
			$\mu_4$ & $-1$ & $-1$ & $ 5$ & $-1$ & $-1$\\
			$\mu_5$ & $-1$ & $-1$ & $-1$ & $ 5$ & $-1$\\
			$\mu_6$ & $-1$ & $-1$ & $-1$ & $-1$ & $11$
		\end{tabular}
		\label{tab_Minkowski sum nabla}
	}
	\caption{\small Lattice polytope $\Delta^\circ$ for the weighted projective space $\IP^5_{112233}$ and its polar polytope $\Delta$.}
\end{table}

Let us consider a specific example to illustrate what we learned so far. Let $X$ be the 5-dimensional weighted projective space $\IP^5_{112233}$, which is described by the polytope $\smash{\Delta^\circ_{\IP^5_{112233}}}$ from table \ref{tab_P112233}. Consider a partition of $\Delta^\circ$ into the subsets $\Delta^\circ_1 = \text{Conv}(\left\{\nu_1,\nu_2,\nu_4\right\},0)$ and $\Delta^\circ_2 = \text{Conv}(\left\{\nu_3,\nu_5,\nu_6\right\},0)$. One can show that $\Delta^\circ_1+\Delta^\circ_2$ is a reflexive polytope and hence this partition is indeed a nef partition. Using \eqref{eq:Cayley polytope} we can calculate $P^\ast_{\Delta^\circ}$, see table \ref{tab:Cayley P112233}, and by employing \eqref{eq_dualize Cayley} the dual Cayley polytope $P_{\Delta^\circ}$ given in table \ref{tab_dual Cayley P112233} follows. From $P_{\Delta^\circ}$ it is quite easy to read off $\nabla$ and its nef partition $\{\nabla_1,\nabla_2\} $ and we can confirm that the Minkowski sum
\beq
  \nabla_1+\nabla_2
\eeq
of this partition indeed equals the polar polytope of $\Delta^\circ$ in table~\ref{tab_Minkowski sum nabla}.

\begin{table}[ht]
	\subfloat[The Cayley polytope]{
		\centering
		\begin{tabular}{r@{\,$=$\,(\,}r@{,\;\;}r@{,\;\;}r@{,\;\;}r@{,\;\;}r@{,\;\;}r@{,\;\;}r@{\,)\;\;}}
			\multicolumn{8}{c}{Vertices of $P^\ast_{\Delta^\circ}$}\\ 
			\hline\hline
			$\tilde\nu_1{}^\big.$ & $-1$ & $-2$ & $-2$ & $-3$ & $-3$ & $~1$ & $~0$\\
			$\tilde\nu_2$ & $ 1$ & $ 0$ & $ 0$ & $ 0$ & $ 0$ & $ 1$ & $ 0$\\
			$\tilde\nu_3$ & $ 0$ & $ 1$ & $ 0$ & $ 0$ & $ 0$ & $ 0$ & $ 1$\\
			$\tilde\nu_4$ & $ 0$ & $ 0$ & $ 1$ & $ 0$ & $ 0$ & $ 1$ & $ 0$\\
			$\tilde\nu_5$ & $ 0$ & $ 0$ & $ 0$ & $ 1$ & $ 0$ & $ 0$ & $ 1$\\
			$\tilde\nu_6$ & $ 0$ & $ 0$ & $ 0$ & $ 0$ & $ 1$ & $ 0$ & $ 1$\\
			$\tilde\nu_7$ & $ 0$ & $ 0$ & $ 0$ & $ 0$ & $ 0$ & $ 1$ & $ 0$\\
			$\tilde\nu_8$ & $ 0$ & $ 0$ & $ 0$ & $ 0$ & $ 0$ & $ 0$ & $ 1$
		\end{tabular}
		\label{tab:Cayley P112233}
	}
	\subfloat[The dual Cayley polytope]{
		\centering
		\begin{tabular}{r@{\,$=$\,(\,}r@{,\;\;}r@{,\;\;}r@{,\;\;}r@{,\;\;}r@{,\;\;}r@{,\;\;}r@{\,)\;\;}}
			\multicolumn{8}{c}{Vertices of $P_{\Delta^\circ}$}\\ 
			\hline\hline
			$\tilde\mu_1{}^\big.$  & $ 0$ & $-1$ & $ 0$ & $-1$ & $-1$ & $~1$ & $~0$\\
			$\tilde\mu_2$  & $ 2$ & $-1$ & $ 0$ & $-1$ & $-1$ & $ 1$ & $ 0$\\
			$\tilde\mu_3$  & $ 0$ & $ 1$ & $ 0$ & $-1$ & $-1$ & $ 1$ & $ 0$\\
			$\tilde\mu_4$  & $ 0$ & $-1$ & $ 3$ & $-1$ & $-1$ & $ 1$ & $ 0$\\
			$\tilde\mu_5$  & $ 0$ & $-1$ & $ 0$ & $ 2$ & $-1$ & $ 1$ & $ 0$\\
			$\tilde\mu_6$  & $ 0$ & $-1$ & $ 0$ & $-1$ & $ 5$ & $ 1$ & $ 0$\\[1.5mm]
			$\tilde\mu_7$  & $-1$ & $ 0$ & $-1$ & $ 0$ & $ 0$ & $ 0$ & $ 1$\\
			$\tilde\mu_8$  & $ 1$ & $ 0$ & $-1$ & $ 0$ & $ 0$ & $ 0$ & $ 1$\\
			$\tilde\mu_9$  & $-1$ & $ 2$ & $-1$ & $ 0$ & $ 0$ & $ 0$ & $ 1$\\
			$\tilde\mu_{10}$ & $-1$ & $ 0$ & $ 2$ & $ 0$ & $ 0$ & $ 0$ & $ 1$\\
			$\tilde\mu_{11}$ & $-1$ & $ 0$ & $-1$ & $ 3$ & $ 0$ & $ 0$ & $ 1$\\
			$\tilde\mu_{12}$ & $-1$ & $ 0$ & $-1$ & $ 0$ & $ 6$ & $ 0$ & $ 1$
		\end{tabular}
		\label{tab_dual Cayley P112233}
	}
	\caption{\small The Cayley Polytope and the dual Cayley polytope corresponding to the nef partition $\Delta^\circ_1 = \text{Conv}(\left\{\nu_1,\nu_2,\nu_4\right\},0)$ and $\Delta^\circ_2 = \text{Conv}(\left\{\nu_3,\nu_5,\nu_6\right\},0)$ of $\IP^5_{112233}$ as well as to the dual nef partition $\nabla_1$ and $\nabla_2$ coming from $\left\{\tilde\mu_1,\dots,\tilde\mu_6\right\}$ and $\left\{\tilde\mu_7,\dots,\tilde\mu_{12}\right\}$, respectively. The nef partition of the dual Cayley polytope can easily be read off from the last two columns.}
\end{table}

\subsubsection*{The stringy $E$-function}
The generalization of equations \eqref{eq:Batyrev1} and \eqref{eq:Batyrev2} to complete intersections will be a formula that counts faces in the Cayley and the dual Cayley polytope instead of the original one and its polar. The formula itself will also become a bit more complicated compared to the one for hypersurfaces.

In \cite{BatBor} Batyrev and Borisov introduced a generating function for the so called stringy Hodge numbers of a CICY corresponding to the introduced Cayley cone above. These stringy Hodge numbers are equal to the usual Hodge numbers in case that a crepant resolution of the generically singular Calabi-Yau exists. They are given as coefficients of the stringy $E$-function, namely
\beq\label{eq_stringy E-function 1}
  E_\text{st}(S;u,v) = \sum_{p,q} (-1)^{p+q} h_\text{st}^{p,q}(S)~u^pv^q.
\eeq
The generalization to arbitrary Gorenstein polytopes was done by Batyrev and Nill in \cite{BatNil}. There it was conjectured that the stringy $E$-function is actually a polynomial in $u,v$ which was proven recently by Nill and Schepers in \cite{NillSchep}. In terms of the dual Cayley polytope, the stringy $E$-function can be expanded as
\beq\label{eq_stringy E-function 2}
  E(u,v)=\frac{1}{(uv)^l}\sum_{{}\quad\,\,\,\mclap{\emptyset\leq x \leq y \leq P}}(-1)^{1+\dim x} u^{1+\dim y} S_x\left(\frac{u}{v}\right) S_{y^\vee}(uv)B_{\left[x,y\right]}(u^{-1},v).
\eeq
Here we sum over faces of $P$. These form an (Eulerian) partially ordered set (poset) where the partial ordering is given by
\beq
  x\leq y \Leftrightarrow x\text{ is a face of }y
\eeq
and denote the poset of all faces of $P$ by $\mathcal P$. Furthermore we define
\beq
  \bal
    \left[x,y\right] &\ce \left\{z\in\mathcal P : x\leq z\leq y\right\}\quad \text{ as well as} \\
    y_C^\vee &\ce \left\{n\in C: \left<m,n\right> = 0,~\forall m\in C^\ast\right\}
	\eal
\eeq
to be the dual face of a face $y_C$ in $C$. Since there is a relation between faces in $C$ and $C^\ast$ we also get such a correspondence for faces in the supporting polytopes $P$ and $P^\ast$. The dual face to $y$ in $P$ is denoted by $y^\vee$ in $P^\ast$.

For some $k$-dimensional face $\mathcal F$, the polynomial $S_\mathcal F(t)$ is defined by
\beq
  S_\mathcal F(t) \ce(1-t)^{k+1} \sum_{i=0}^\infty l(k\mathcal F)\,  t^i\,.
\eeq
For the last missing piece, let $\mathcal P'$ be some subposet of $\mathcal P$ with minimal element $\hat 0$, maximal element $\hat 1$ and dim$(\hat 1)=k-1$. The Batyrev-Borisov polynomials $B_{\mathcal P'} (u,v)$ are defined recursively in terms of the polynomials $H_{\mathcal P'}(t)$ and $G_{\mathcal P'}(t)$. We set $H_{\mathcal P'}(t)=G_{\mathcal P'}(t)=1$ if $k=0$ and for $k>0$ we define
\beq
  \bal
    H_{\mathcal P'}(t)&\ce\sum_{\mclap{\hat 0<x<\hat 1}} (t-1)^{\dim(x)} G_{[x,\hat 1]}(t),\\
    G_{\mathcal P'}(t)&\ce\tau_{<\frac{k}{2}}(1-t) H_{\mathcal P'}(t),
	\eal
\eeq
where $\tau_{<\frac{k}{2}}$ is the truncation operator defined by its action on sums
\beq
  \tau_{<\frac{k}{2}} \sum_{i=0}^\infty a_i t^i \ce \sum_{\mclap{0\leq i < \frac k 2}} a_it^i .
\eeq
The Batyrev-Borisov polynomials start also with $B_{\mathcal P'} (u,v)=1$ for $k=0$ and can be read off from the following formula for $k>0$: 
\beq
  \sum_{\mclap{\hat 0<x<\hat 1}} B_{[\hat 0,x]}(u,v)u^{k-\dim(x)+1}G_{[x,\hat
      1]}(u^{-1}v)\ce G_{\mathcal P'}(uv)\; .
\eeq

\subsubsection*{Closed form expressions for $h^{1,1}$ and $h^{d-3,1}$ of a CICY}
The above formula \eqref{eq_stringy E-function 2} is neither particularly elegant nor easy to work with. Using some simplifying relations for the polynomials above, stated for instance in \cite{BatBor} or alternatively in \cite{Novoseltsev} propositions 2.2 and 2.4, one can  deduce the formulas \eqref{eq:Batyrev1} and \eqref{eq:Batyrev2} from \eqref{eq_stringy E-function 2} (see Theorem 3.1 in \cite{Novoseltsev}). Similarly in the same paper Doran and Novoseltsev deduced an explicit closed form expression for the Hodge numbers of Calabi-Yau varieties realized as a complete intersection of two hypersurfaces in a 5-dimensional toric variety. 

To present their result, we first recall that the nef partition is called indecomposable, if no subset of $\left\{\Delta^\circ_i,i=1,\dots,l\right\}$ exists whose Minkowski sum is reflexive, namely if we are not dealing with a product of Calabi-Yau manifolds. If this is the case, the combinatorial formula for the Hodge numbers is given as follows:
\beq\label{eq:Novoseltsev1}
	\bal
		h^{1,1}(S) = {}&{} l(P^\ast)-7 \\
		 &{} -  \sum_{\mclap{\dim(y)=0}} l^\ast(2y^\vee) +  \sum_{\mclap{\dim(y)=1}} l^\ast(y^\vee) \\
		&{}+ \sum_{\mclap{\dim(y)=1}} l^\ast(y)\cdot l^\ast(2y^\vee) - \sum_{\mclap{\dim(y)=2}} l^\ast(2y)\cdot l^\ast(y^\vee) \\
		&{}-\sum_{\mclap{\substack{\dim(x)=2\\\dim(y)=3\\x<y}}} l^\ast(x)\cdot l^\ast(y^\vee) + \sum_{\mclap{\dim(y)=3}} l^\ast(2y)\cdot l^\ast(y^\vee).
	\eal
\eeq
By taking all dual polytopes and faces we obtain of course the second Hodge number as
\beq\label{eq:Novoseltsev2}
	\bal
		h^{2,1}(S) ={}&{} l(P)-7 - \sum_{\mclap{\dim(y_\ast)=0}} l^\ast(2y_\ast^\vee) + \sum_{\mclap{\dim(y_\ast)=1}} l^\ast(y_\ast^\vee) \\
		&{} + \sum_{\mclap{\dim(y_\ast)=1}} l^\ast(y_\ast)\cdot l^\ast(2y_\ast^\vee) - \sum_{\mclap{\dim(y_\ast)=2}} l^\ast(2y_\ast)\cdot l^\ast(y_\ast^\vee) \\
		&{} - \sum_{\mclap{\substack{\dim(x_\ast)=2\\\dim(y_\ast)=3\\x_\ast<y_\ast}}} l^\ast(x_\ast)\cdot l^\ast(y_\ast^\vee) + \sum_{\mclap{\dim(y_\ast)=3}} l^\ast(2y_\ast)\cdot l^\ast(y_\ast^\vee).
	\eal
\eeq
Here $x,y$ and $x_\ast,y_\ast$ are faces of $P$ and $P_\ast$, respectively. 

Having  a close look at the proof of formulas \eqref{eq:Novoseltsev1} and \eqref{eq:Novoseltsev2}, one realizes that it is not too hard to generalize it to the case of two complete intersections in a 6-dimensional ambient space, yielding a Calabi-Yau 4-fold. Due to recent interest in such 4-folds in the context of F-theory we deduced a closed form expression for such cases at least for the Hodge numbers $h^{1,1}$ and $h^{3,1}$ via the dual computation. The result is in the most generic form and no simplifications have been taken into account. For a Calabi-Yau 4-fold $S$ we find:
\beq\label{eq:werder1}
  \bal
		h^{1,1}(S) ={}&{} l(P^\ast)-8 - \sum_{\mclap{\dim(y)=0}}\left[l^\ast(2y^\vee)-7l^\ast(y^\vee)\right] \\
		&{} + \sum_{\mclap{\dim(y)=1}} l^\ast(y^\vee) - \sum_{\mclap{\dim(y)=2}} \left[ k(y) -3  \right]\cdot l^\ast(y^\vee) \\
		&{} + \sum_{\mclap{\dim(y)=1}} l^\ast(y)\cdot\left[l^\ast(2y)-6\cdot l^\ast(y^\vee)\right] - \sum_{\mclap{\substack{\dim(x)=1\\\dim(y)=2\\x<y}}} l^\ast(x)\cdot l^\ast(y^\vee)\\
		&{} - \sum_{\mclap{\substack{\dim(x)=2\\\dim(y)=3\\x<y}}} l^\ast(x)\cdot l^\ast(y^\vee) + \sum_{\mclap{\dim(y)=3}} l^\ast(y^\vee)\cdot\left[l^\ast(2y)-4l^\ast(y)\right]\,,
  \eal
\eeq
along with its mirror dual Hodge number
\beq
\label{eq:werder2}
  \bal
		h^{3,1}(S) ={}&{} l(P)-8 - \sum_{\mclap{\dim(y_\ast)=0}} \left[l^\ast(2y_\ast^\vee)-7l^\ast(y_\ast^\vee)\right] \\
		&{} + \sum_{\mclap{\dim(y_\ast)=1}} l^\ast(y_\ast^\vee) - \sum_{\mclap{\dim(y_\ast)=2}} \left[ k(y_\ast) -3  \right]\cdot l^\ast(y_\ast^\vee) \\
		&{} + \sum_{\mclap{\dim(y_\ast)=1}} l^\ast(y_\ast)\cdot\left[l^\ast(2y_\ast)-6\cdot l^\ast(y_\ast^\vee)\right] - \sum_{\mclap{\substack{\dim(x_\ast)=1\\\dim(y_\ast)=2\\x_\ast<y_\ast}}} l^\ast(x_\ast)\cdot l^\ast(y_\ast^\vee)\\
		&{} - \sum_{\mclap{\substack{\dim(x_\ast)=2\\\dim(y_\ast)=3\\x_\ast<y_\ast}}} l^\ast(x_\ast)\cdot l^\ast(y_\ast^\vee) + \sum_{\mclap{\dim(y_\ast)=3}} l^\ast(y_\ast^\vee)\cdot\left[l^\ast(2y_\ast)-4l^\ast(y_\ast)\right]\,.
	\eal
\eeq
Here $k(y)$ denotes the number of vertices of the face $y$. It is quite nice to have such an explicit formula at hand but one has also to admit that the actual calculation is not that easy to perform and may take some time. Our implementation in Macaulay2 using the {\tt Polyhedra} package \cite{Polyhedral.m2} needed at least ten minutes to produce the numbers. Using {\cohomCalg} \cite{cohomCalg:Implementation} we were able to obtain the same results along with the remaining two Hodge numbers much faster for varieties where $h^{1,1}(S)$ is not too large.

\subsubsection*{Correspondence to line bundle cohomologies of CICY}
The question now is, whether also for these CICYs one can identify terms in the generalized Bartyrev formulas with line bundle cohomology classes of the ambient five-fold respectively six-fold.

Let us here restrict to the case of a CICY in an ambient five-fold, i.e.~the formulas \eqref{eq:Novoseltsev1} and \eqref{eq:Novoseltsev2}. Indeed, we observed that, if we calculate those Hodge numbers via exact sequences and therefore via line bundle cohomologies of the ambient toric variety --- using {\cohomCalg} \cite{cohomCalg:Implementation} as described in \ref{subsection:Complete intersection (CI) subvarieties} --- we find that the following relation between the terms in the combinatorial formula for the $h^{1,1}(S)$ and the line bundle cohomologies applies:
\beq
	\boxed{
		\bal
			\fhh^{1,1}_5(S)  &{}= l(P^\ast)-7\,,\\
			\fhh^{1,1}_4(S)  &{}= -  \sum_{\mclap{\dim(y)=0}} l^\ast(2y^\vee) +  \sum_{\mclap{\dim(y)=1}} l^\ast(y^\vee), \\
			 \fhh^{1,1}_3(S)  &{}= \sum_{\mclap{\dim(y)=1}} l^\ast(y)\cdot l^\ast(2y^\vee) - \sum_{\mclap{\dim(y)=2}} l^\ast(2y)\cdot l^\ast(y^\vee), \\
			\fhh^{1,1}_2(S)  &{}= - \sum_{\mclap{\substack{\dim(x)=2\\\dim(y)=3\\x<y}}} l^\ast(x)\cdot l^\ast(y^\vee) + \sum_{\mclap{\dim(y)=3}} l^\ast(2y)\cdot l^\ast(y^\vee) 
		\eal
	}
\eeq
Here $\fhh^{p,q}_i(S)$ are defined as in section \ref{subsection:Complete intersection (CI) subvarieties} equation \eqref{eq_weirdcontributionnumbers}. Similarly, we find the relations for $h^{2,1}(S)$ Hodge number to be
\beq
	\boxed{
		\bal
			\fhh^{2,1}_0(S) &{} = l(P)-7 - \sum_{\mclap{\dim(y_\ast)=0}} l^\ast(2y_\ast^\vee) + \sum_{\mclap{\dim(y_\ast)=1}} l^\ast(y_\ast^\vee) \\
			\fhh^{2,1}_1(S) &{}= \sum_{\mclap{\dim(y_\ast)=1}} l^\ast(y_\ast)\cdot l^\ast(2y_\ast^\vee) - \sum_{\mclap{\dim(y_\ast)=2}} l^\ast(2y_\ast)\cdot l^\ast(y_\ast^\vee) \\
			\fhh^{2,1}_2(S) &{}= -\sum_{\mclap{\substack{\dim(x_\ast)=2\\\dim(y_\ast)=3\\x_\ast<y_\ast}}} l^\ast(x_\ast)\cdot l^\ast(y_\ast^\vee) + \sum_{\mclap{\dim(y_\ast)=3}} l^\ast(2y_\ast)\cdot l^\ast(y_\ast^\vee)
		\eal
	}
\eeq
For $h^{2,1}(S)$, as in \eqref{eq:Batyrev2}, we note that terms where no mixing of faces and dual faces takes place correspond to global sections in line bundles whereas terms where such a mixing does happen correspond to higher cohomology classes of line bundles of $X$. In \eqref{eq:Novoseltsev2} for 1 hypersurface we got at most $h^1_X(\cdot)$ contributions, where now in case of 2 hypersurfaces we found at most $h^2_X(\cdot)$ contributions to the complex structure moduli.

Since the computations get quite involved, we have not yet identified the analogous correspondence for  the case of a CICY in a toric six-fold, i.e.~eq.~\eqref{eq:werder1} and \eqref{eq:werder2}, but have no doubt that it exists and takes a similar form.

\section{Summary and Conclusions}
In this paper we have discussed many applications of our algorithm for the computation of line bundle valued cohomology over toric varieties. Specifically, we considered cohomological questions arising in compactifications of the heterotic and the Type~II superstring to four dimensions. We collected the necessary material and topological tools in this single paper and hope that it serves as a guide both for string model building in the geometric phase and for the computational diversity of our algorithm.

The main new result is the generalization of the algorithm to equivariant cohomology, which is important for the study of orientifolds and orbifolds. Moreover, it was possible to give the various terms in the Batyrev (like) formulas a clear cohomological interpretation. In particular, the so-called twisted contributions could be identified with higher cohomology classes.

\subsection*{Acknowledgment}
We gratefully thank our computers for doing most of the work for us. The authors would like to thank Andr\'{e}s Collinucci for some helpful discussions. T.~Rahn and H.~Roschy would like to thank Ren\'e Birkner for providing us with some supplemental routines to his Macaulay2 package {\tt Polyhedra} which we used to calculate various examples. R.~Blumenhagen and B.~Jurke would like to thank the Kavli Institute for Theoretical Physics, Santa Barbara for the hospitality during the early stages of the project. We also would like to thank Andreas Deser for some helpful comments on the manuscript. This research was supported in part by the National Science Foundation under Grant No.~PHY05-51164.

\appendix

\section{Counting bundle moduli}

In heterotic constructions one often needs to count the number of bundle deformation moduli, which are counted by the number of distinct endomorphisms a bundle supports. Therefore we need to count the number of global sections of ${\rm End}(V) \cong V \otimes V^*$. In this appendix we are going to show that this problem can --- at least in principle --- also be addressed using the methods presented in the main part of this paper. Like for the examples in section~\ref{sec_calabiyau} we approach this in two steps: first we are considering the situation on the ambient space and then via the Koszul sequence pull it down to a hypersurface.

\subsubsection*{Ambient space}
In the case of a monad bundle $V$ defined via \eqref{eq_monadbundleprototype}, one can work out the corresponding sequences explicitly. Let us use the generic abbreviation
\beq
  \cO(\vec a) \ce \bigoplus_{i=1}^{\mclap{\dim \vec a}} \cO_X(a_i),
\eeq
then we obtain from tensoring the monad bundle short exact sequence \eqref{eq_monadbundleprototype} with the dual vector bundle $V^*$ the sequence
\beq
  0 \fto V \otimes V^* \injto \cO(\vec b)\otimes V^* \surjto \cO(\vec c)\otimes V^* \fto 0.
\eeq
In order to determine $\cO(\vec b)\otimes V^*$ and $\cO(\vec c)\otimes V^*$ we tensor the dual sequence of \eqref{eq_monadbundleprototype} --- which is itself of the monad bundle structure \eqref{eq_dualmonadbundleprototype} --- with $\smash{\cO(\vec b)}$ and $\smash{\cO(\vec c)}$, respectively, i.e.
\beq
  \bal
    & 0 \fto \cO(\vec b) \otimes \cO(\vec c)^* \injto \cO(\vec b) \otimes \cO(\vec b)^* \surjto \cO(\vec b)\otimes V^* \fto 0 \\
    & 0 \fto \cO(\vec c) \otimes \cO(\vec c)^* \injto \cO(\vec c) \otimes \cO(\vec b)^* \surjto \cO(\vec c)\otimes V^* \fto 0
  \eal
\eeq

We can proceed completely analogous for the extension bundle construction, where the endomorphism bundle is obtained from
\beq
  0 \fto \cO(\vec a)\otimes W^* \injto W\otimes W^* \surjto \cO(\vec c)\otimes W^* \fto 0
\eeq
and the two required products $\cO(\vec a) \otimes W^*$ and $\cO(\vec c)\otimes W^*$ can be obtained from
\beq
  \bal
    & 0 \fto \cO(\vec a)\otimes\cO(\vec c)^* \injto \cO(\vec a)\otimes W^* \surjto \cO(\vec a)\otimes\cO(\vec a)^* \fto 0 \\
    & 0 \fto \cO(\vec c)\otimes\cO(\vec c)^* \injto \cO(\vec c)\otimes W^* \surjto \cO(\vec c)\otimes\cO(\vec a)^* \fto 0.
  \eal
\eeq

Ultimately, the resulting sequences for the endomorphism bundle of a monad bundle construction \eqref{eq_monadbundleprototype} are then
\beq
  \bal
    & 0 \fto {\rm End}(V) \injto \cO(\vec b)\otimes V^* \surjto \cO(\vec c)\otimes V^* \fto 0 \\
    & 0 \fto \bigoplus_{i=1}^{r_B} \bigoplus_{j=1}^{r_C} \cO_X(b_i-c_j) \injto \bigoplus_{i=1}^{r_B} \bigoplus_{j=1}^{r_B} \cO_X(b_i-b_j) \surjto \cO(\vec b)\otimes V^* \fto 0 \\
    & 0 \fto \bigoplus_{i=1}^{r_C} \bigoplus_{j=1}^{r_C} \cO_X(c_i-c_j) \injto \bigoplus_{i=1}^{r_C} \bigoplus_{j=1}^{r_B} \cO_X(c_i-b_j) \surjto \cO(\vec c)\otimes V^* \fto 0
  \eal
\eeq
and likewise for the extension bundles \eqref{eq_extensionsbundleprototype}
\beq
  \bal
    & 0 \fto \cO(\vec a)\otimes W^* \injto {\rm End}(W) \surjto \cO(\vec c)\otimes W^* \fto 0 \\
    & 0 \fto \bigoplus_{i=1}^{r_A} \bigoplus_{j=1}^{r_C} \cO_X(a_i-c_j) \injto \cO(\vec a)\otimes W^* \surjto \bigoplus_{i=1}^{r_A} \bigoplus_{j=1}^{r_A} \cO_X(a_i-a_j) \fto 0 \\
    & 0 \fto \bigoplus_{i=1}^{r_C} \bigoplus_{j=1}^{r_C} \cO_X(c_i-c_j) \injto \cO(\vec c)\otimes W^* \surjto \bigoplus_{i=1}^{r_C} \bigoplus_{j=1}^{r_A} \cO_X(c_i-a_j) \fto 0 \\
  \eal
\eeq
In order to compute $h^\bullet(X;{\rm End}(V))$ or $h^\bullet(X;{\rm End}(W))$ it is therefore necessary to evaluate two auxiliary bundle cohomologies.

\subsubsection*{Hypersurfaces}
In case a hypersurface is considered, the general approach is the same, however a lot more sequences are involved, as one has to employ a splitting of the monad bundle sequence analogous to the splitting of the tangent bundle sequence. From the second variant of the monad bundle construction \eqref{eq_dualmonadbundleprototype} we therefore obtain
\beq
  \bal
    & 0 \fto \cO_D(\vec a) \injto \cO_D(\vec b) \surjto \cE_D \fto 0 \\
    & 0 \fto V_D \injto \cE_D \surjto \cO_D(D) \fto 0,
  \eal
\eeq
where $V_D$ is the monad bundle on the hypersurface $D$, $\cE_D$ is the auxiliary bundle from the split sequence and a lot of (neglected) twisted Koszul sequences are involved in order to construct the individual line bundles $\cO_D(a_i)$ of $\cO_D(\vec a)$ on the hypersurface. Following the same idea as before, the endomorphism bundle is obtained by tensoring with the dual bundle $V_D^*$ which gives
\beq
  0 \fto {\rm End}(V_D) \injto \cE_D\otimes V_D^* \surjto \cO_D(D)\otimes V_D^* \fto 0.
\eeq
In order to compute $\cE_D\otimes V_D^*$ we need to consider the sequences
\beq
  \bal
    & 0 \fto \cO_D(\vec a) \otimes V_D^* \injto \cO_D(\vec b)\otimes V_D^* \surjto \cE_D\otimes V_D^* \fto 0 \\
    & 0 \fto \cO_D(\vec a) \otimes \cO_D(D)^* \injto \cO_D(\vec a)\otimes \cE_D^* \surjto \cO_D(\vec a) \otimes V_D^* \fto 0 \\
    & 0 \fto \cO_D(\vec a) \otimes \cE_D^* \injto \cO_D(\vec a) \otimes \cO_D(\vec b)^* \surjto \cO_D(\vec a)\otimes \cO_D(\vec a)^* \fto 0 \\
    & 0 \fto \cO_D(\vec b) \otimes \cO_D(D)^* \injto \cO_D(\vec b)\otimes \cE_D^* \surjto \cO_D(\vec b) \otimes V_D^* \fto 0 \\
    & 0 \fto \cO_D(\vec b) \otimes \cE_D^* \injto \cO_D(\vec b) \otimes \cO_D(\vec b)^* \surjto \cO_D(\vec b)\otimes \cO_D(\vec a)^* \fto 0
  \eal
\eeq
and for the $D$-twisted dual monad bundle $\cO_D(D)\otimes V_D^*$ we have to compute
\beq
  \bal
    & 0 \fto \cO_D(D)\otimes\cO_D(D)^* \injto \cO_D(D)\otimes\cE_D^* \surjto \cO_D(D)\otimes V_D^* \fto 0 \\
    & 0 \fto \cO_D(D) \otimes \cE_D^* \injto \cO_D(D)\otimes\cO_D(\vec b)^* \surjto \cO_D(D) \otimes \cO_D(\vec a)^* \fto 0
  \eal
\eeq

As one can see from those sequences, the general structure remains fairly simple, however, it becomes rather laborious to work through all the sequences. It should be mentioned that the ``evaluation via exactness'', which has worked quite well so far, often fails in the evaluation of those endomorphism bundle sequences. One simply does not get the required number of zeros in the involved intermediate cohomologies, which requires to actually compute the induced cohomology mappings. One can also consider entirely different methods, e.g.~in \cite{Aspinwall:2010ve} the tangent bundle deformation moduli are computed via spectral sequences.


\clearpage
\bibliography{rev23}
\bibliographystyle{utphys}

\end{document}